\documentclass[fleqn,usenatbib]{mnras}

\usepackage{natbib}
\usepackage[caption=false]{subfig}    
\usepackage{graphicx}
\usepackage{appendix}
\usepackage{amsmath}
\usepackage{amssymb}
\usepackage{color}
\usepackage{tablefootnote}

\newcommand{\eg}{e.g.,}
\newcommand{\ie}{i.e.,}

\newcommand{\etal}{et~al.}

\newcommand{\logten}{\ensuremath{\log_{10}\,}}
\usepackage{cleveref}
\crefname{section}{\S}{\S\S}
\Crefname{section}{\S}{\S\S}

\title[]{Studying the Physical Properties of Tidal Features I. Extracting Morphological Substructure in CANDELS Observations and VELA Simulations}
\author[K. B. Mantha et al.]{Kameswara Bharadwaj Mantha$^{1}$\thanks{E-mail: km4n6@mail.umkc.edu}, Daniel H. McIntosh$^{1}$, Cody P. Ciaschi$^{1}$,
	\newauthor   Rubyet Evan$^{1}$, Henry C. Ferguson$^{2}$, Logan B. Fries$^{1}$,  Yicheng Guo$^{3}$, 
	\newauthor Anton M. Koekemoer$^{2}$, Luther D. Landry$^{1}$, Elizabeth J. McGrath$^{4}$,
	\newauthor Raymond C. Simons$^{5}$,  Gregory F. Snyder$^{2}$, Scott E. Thompson$^{1}$,
	\newauthor Eric F. Bell$^{6}$, Daniel Ceverino$^{7}$,  Nimish P. Hathi$^{2}$,  Camilla Pacifici$^{2}$, 
	\newauthor Joel R. Primack$^{8,9}$, Marc Rafelski$^{2}$, Vicente Rodriguez-Gomez$^{5}$.
	\\
	Affiliations are listed at the end of this paper\\
}
\date{Accepted 2019 March 19. Received 2019 March 16; in original form 2018 October 19.}

\pubyear{2017}

\begin{document}
	\maketitle


	\begin{abstract}
	{The role of major mergers in galaxy evolution remains a key open question. Existing empirical merger identification methods use non-parametric and subjective visual classifications which can pose systematic challenges to constraining merger histories. As a first step towards overcoming these challenges, we develop and share publicly a new Python-based software tool that identifies and extracts the flux-wise and area-wise significant contiguous regions from the model-subtracted “residual” images produced by popular parametric light-profile fitting tools (e.g., GALFIT). Using Hubble Space Telescope ({\it HST}) $H$-band single-S\'ersic residual images of $17$ CANDELS galaxies, we demonstrate the tool’s ability to measure the surface brightness and improve the qualitative identification of a variety of common residual features (disk structures, spiral substructures, plausible tidal features, and strong gravitational arcs). We test our method on synthetic {\it HST} observations of a $z\sim 1.5$ major merger from the VELA hydrodynamic simulations. We extract {\it H}-band residual features corresponding to the birth, growth, and fading of tidal features during different stages and viewing orientations at CANDELS depths and resolution. We find that the extracted features at shallow depths have noisy visual appearance and are susceptible to viewing angle effects. For a VELA $z\sim3$ major merger, we find that James Webb Space Telescope NIRCam observations can probe high-redshift tidal features with considerable advantage over existing {\it HST} capabilities. Further quantitative analysis of plausible tidal features extracted with our new software hold promise for the robust identification of hallmark merger signatures and corresponding improvements to merger rate constraints.}
		
	\end{abstract}
	
	\section{Motivation}
	Merging of two similar-mass (stellar-mass ratio $\leq4:1$) galaxies is often referred to in the literature as major merging. Identifying and quantifying the rate at which galaxies experience such mergers (merger rates) over cosmic history is a key step to empirically quantify the role of major mergers in galaxy evolution. Conceptually, two broad observational approaches are employed to quantify major merging, namely: (1) {\it close-pair methods}  and (2) {\it morphological methods}. These approaches yield redshift-evolutionary trends that broadly agree with theoretical merger rate predictions that major mergers become more frequent with increasing redshift out to $z<1.5$. However, both methodologies suffer from unique, yet analogous systematic biases from selection effects and uncertain observability timescale assumptions, making it difficult to interpret the merger rates at $z>1.5$. \cite{lotz_effect_2010} investigated the effects of such systematics in binary merger simulations using non-parametric identifiers of plausible merging (\eg\, Gini-M$_{20}$ and $CAS$ metrics). In this study, we attempt to help overcome the subjective identification of `hallmark' merger signatures (\eg\, tidal arms, tails, bridges, and extended fans) and to robustly quantify their observability timescales  -- the largest sources of uncertainty towards merger rate estimation in morphology-based methods -- by developing a new {public {\tt Python}} analysis tool to {\it extract and quantify} the morphological substructure of galaxies. When applied to both empirical and simulated data, this tool will facilitate future efforts to calibrate the observables associated with galaxy merging signatures and lead to improved merger rate constraints.
	
	Substantial efforts have attempted to constrain major merger rates as a function of cosmic time using empirical ``merger fractions'' based on close-pair statistics, typically identifying galaxy-galaxy pairs in close physical proximity {\citep[\eg ][]{Mundy17,Mantha18}} or by measuring clustering statistics \citep[\eg][]{bell06b,robaina_merger-driven_2010}. This method leverages on the long-standing idea (supported by numerical expectations) that galaxies in close proximity become gravitationally bound and merge into a more-massive system \citep{Toomre77,Barnes88,Carlberg94,Patton00,Kitzbichler_White_08}. Many previous close-pair-based works broadly agree that merger rates grow more frequent at earlier cosmic times during the redshift range $0\lesssim z\lesssim1.5$, albeit with a wide range of redshift dependencies $\propto(1+z)^{0.5-3}$ {\protect\citep[\eg\,][]{Zepf89,Patton97,Lin04,kartaltepe_evolution_2007,hsieh08,lin_redshift_2008,de_ravel09,de_ravel_zcosmos_2011,Robotham14}}. Such variance is largely caused by different study-to-study close-pair selection assumptions \citep{lotz_major_2011}. In \citet{Mantha18}, we systematically quantified the effect of different close pair selection choices on the derived merger fractions.
	
	Recent efforts extend the close pair method to higher redshifts and find that merger rates may be un-evolving or diminishing with increasing redshift between $1.5\leq z \leq 3$ \citep{Ryan08,man16,Mundy17,Mantha18}. These measurements disagree with theoretical predictions of steadily increasing merger rates from $z=0$ to $z=3$ \citep[\eg\,][]{hopkins_mergers_2010,rodriguez-gomez_merger_2015}. Leveraging recent simulation-based work by \citet{snyder17}, observational studies argue that the data-theory discrepancy at $z>1.5$ may be due to the assumption of simplistic (non-evolving) close-pair observability timescale employed in merger rate calculations \citep[][Duncan \etal\,in prep]{Ventou17,Mantha18}. What is needed is a thorough analysis of the observability timescale evolution for both close-pair and morphological methods. In a related study (Mantha et al. in prep), we are constraining the close-pair observability timescale as a function of different pair selection variables. The residual feature extraction tool that we introduce in this paper will aid new efforts to constrain the morphological feature observability timescales.
	
	Besides merger rates derived from close pairs, rates based on morphology have also been measured motivated by simulations showing that merging galaxies exhibit morphologically-disturbed appearance {\citep[e.g.,][]{BarnesHernquist96,Bournaud06,Peirani10}}. These merger constraints can be broadly categorized into rates based on visual identification of disturbed morphologies \citep[\eg\,][]{Wolf05,bell_dry_2006,Jogee09} and {quantitative metrics of large-scale galaxy asymmetries or morphology such as $CAS$ \citep[first introduced by ][]{Abraham96b,Abraham96a} and used by  \cite{conselice03a,conselice_structures_2008,Lopez-Sanjuan09}, and Gini-M$_{20}$ \citep[introduced in][]{Abraham03} and applied extensively by \cite{lotz08,conselice_structures_2009}}. Early morphology-based studies find a broad range of merger rate evolutions $R\propto(1+z)^{2-5}$ at $z\lesssim 1.5$ \citep[see][]{lotz_major_2011}, albeit with significant study-to-study scatter where some studies find no redshift dependence of merger rates \citep[\eg\,][]{cassata05,lotz08}, and sometimes finding up to $25\%-50\%$ of their samples to be merging \citep{conselice03a}. These studies suffer from low-number statistics due to small-volume pencil-beam surveys, redshift-dependent systematic biases induced by cosmological surface brightness dimming and strong morphological {\it k}-corrections when using rest-frame UV images, which can impact the purity and completeness of merger selection. Although the advent of large Hubble Space Telescope ({\it HST}) surveys like CANDELS \citep{Koekemoer,grogin_candels:_2011} alleviate some issues, the observability timescales for morphology-based methods are highly uncertain and may be varying with redshift.

	A possible solution to overcome the systematic, purity, and timescale related challenges faced by the aforementioned morphology-based approaches is to employ more focused morphological signatures as merger identifiers. For example, \citet{Lackner14} used the presence of multiple-nuclei separated by $\lesssim8\,{\rm kpc}$ to select merging galaxies during a narrow window straddling the early and post-merger stages \citep[few hundred Myr timescale;][]{lotz_major_2011}, while aiming to minimize the contamination from spurious non-merging interlopers. Tidal features are a specific sub-class of morphological substructures that are predicted to be prevalent among merging galaxies \citep[\eg][]{Toomre72,Eneev73,BarnesHernquist96}. A fruitful technique for improving the detection of faint and transient tidal features is through the analysis of residual images \footnote{Produced by subtracting the best-fit light profile of the host galaxy using tools such as {\tt GALFIT} \citep{Peng02}.} either by visual inspection \citep[\eg\,][]{bell_dry_2006,mcintosh_2008} or quantitative parameters \citep[\eg\,][]{Tal09,Hoyos12}. In fact, galaxy light-profile fitting is a popular step employed in many large surveys to provide key insights into the structural (size, shape) evolution of galaxies \citep[\eg][]{newman_can_2012,van_der_wel_12,van_der_wel_3d-hst+candels:_2014}. In this context, our novel tool is designed to analyze the by-product residual images from galaxy image fitting routines and extract useful information about additional (often complex) substructures.

	Detailed objective distinction of {\it hallmark} tidal features from other non-merging signatures (\eg\, lopsided disks, asymmetric or one-armed spiral structures) is yet to be achieved. Existing empirical merger-identification methods only provide a {\it plausible} indication of tidal signatures and do not attempt to quantitatively capture key information such as structure (length, width, and pitch), morphology (\eg\, tails and fans), color, and stellar mass of the features. Furthermore, constraining the role of observational effects (\eg\, depth, cosmological surface-brightness dimming) and observability timescales $T_{\rm obs}$ for {\it hallmark} tidal features is an open question, and is a key hurdle in achieving robust tidal-feature-based merger rates.
	
	As a first step towards quantitatively identifying and extracting tidal signatures of galaxy merging, in this study, we introduce a new technique {and its corresponding public {\tt Python} software pipeline\footnote{The data products discussed in this paper along with the software are publicly available at \url{https://github.com/AgentM-GEG/residual_feature_extraction}}} to extract and quantify  {\it residual substructures} and we demonstrate its application with a select sample of 16 CANDELS galaxies that exhibit a range of merging and non-merging morphological substructures. We also test our technique's ability to identify plausible interaction signatures by applying it to mock Hubble Space Telescope ({\it HST}) observations of a galaxy merger from the VELA zoom-in hydrodynamic simulations \citep{Ceverino14,zolotov15}. We structure this paper as follows: In \cref{data}, we introduce our CANDELS test galaxy sample and their GALFIT-derived data products, followed by a brief description of a galaxy merger from the VELA hydro-dynamic simulation and the generation of its mock {\it HST} observations.  In \cref{methodology}, we describe the step-by-step methodology of our residual substructure analysis pipeline, and we apply this method to the select CANDELS galaxies in \cref{extract_res_struc_can}. In \cref{extract_tf_vela}, we apply our methodology to mock observations of a merger simulation and investigate the role of image depth and viewing angle on the feature extraction process. In \cref{discussion}, we present possible applications and limitations of our method. We present our conclusions in \cref{conclusions}. Throughout this work, we adopt the \cite{ade15} cosmological framework ($H_{0}=67.7\,{\rm km\,s^{-1}\,Mpc^{-1}}$), and use the AB magnitude system \citep{oke83}.

	\begin{figure*}
		\includegraphics[width=2\columnwidth]{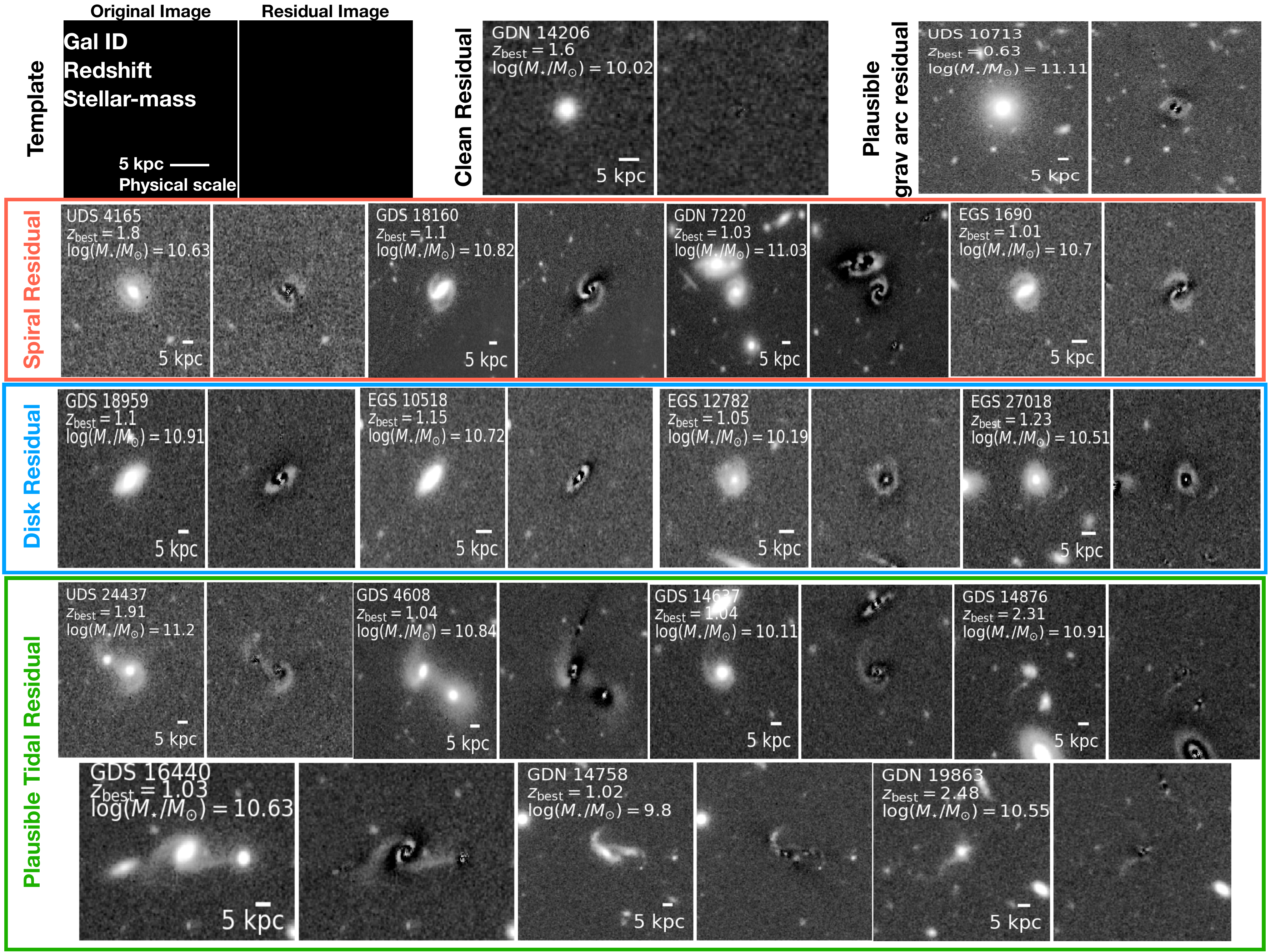}\\
		\caption{Visualization of our demonstrative galaxy sample (\cref{candels_data}, Table\,\ref{table_can_test_sample}) hosting different residual features -- Clean (\ie\, consistent with the background; {\it top-row, middle panel}), plausible gravitational arc residual from {\protect \cite{Hocking18}} ({\it top row, right panel}), spiral arms ({\it second row}), left-over disk ({\it third row}), and plausible interaction signatures ({\it forth row}). We also visually illustrate each panel with a template, where for each galaxy, we show the $H$-band (F160W) image (left) and its corresponding single-S\'ersic residual (\ie\,model-subtracted) image from vdW12 (right). We also report the official CANDELS identification number, the $5\,{\rm kpc}$ physical scale at the galaxy's best-available redshift ($z_{\rm best}$), and stellar mass. }
		\label{clean_pTF_sTF}
	\end{figure*}

	\section{Data and Test Samples}
	\label{data}
	To develop and demonstrate our novel method for extracting and quantifying residual substructures, we use {\it HST}/WFC3 $H$-band (F160W) images and single-S\'ersic {\tt GALFIT} residual images from \cite{van_der_wel_12} for a select sample of galaxies from the {CANDELS survey \citep{grogin_candels:_2011,Koekemoer}} that exhibit different residual substructures including plausible tidal signatures. To explore the application of our method for repeatable identification and extraction of bonafide tidal features, we also analyze mock images derived from a merging galaxy from the VELA zoom-in simulations \citep{Ceverino14}.
	\subsection{CANDELS Data}
	\label{candels_data}
	We make use of the $H$-band S\'ersic-fitting data products from \citealp{van_der_wel_12} (hereafter vdW12) for galaxies bright than $H=24.5\,{\rm mag}$ in the CANDELS survey. Briefly, vdW12 used GALFIT \citep{Peng02} to generate an image-cube containing the {\it original} image used for fitting, the best-fit {\it model} image with the S\'ersic model information stored in its header, and the {\it residual} (original$-$model) image. Following standard procedures outlined in \citet{Barden12}, vdW12 used \citep[{\tt SExtractor};][]{bertin96} to extract sources from the CANDELS imaging \citep{Koekemoer}, estimated the background sky, and determined GALFIT parameter input values, and produced a rectangular image cutout of each galaxy based on its SExtractor parameters: $x,y$ centroid, long side length of five (5) times the Kron radius, and orientated by the position angle $\theta$. In practice, the postage stamp images typically have sides $20-30$ times the galaxy's effective size ($r_{\rm e}$). To ensure optimal galaxy profile fitting with GALFIT, vdW12 masked neighboring objects 4\,mag fainter than the central galaxy and simultaneously fit all remaining objects in the image cutout.
	
	The methodology we describe in \cref{methodology} starts with the vdW12 image cubes for a select sample of $16$ galaxies from an ongoing effort (HST-AR 15040; PI: McIntosh) to visually characterize the $H$-band residual substructures hosted by CANDELS galaxies with $ 9.8 \lesssim \logten(M_{\rm stellar}/M_{\odot}) \lesssim11$ and $0.5\lesssim z \lesssim 2.5$. 
	As shown in Table\,\ref{table_can_test_sample}, these galaxies sample different kinds of residual features: disk ($4$ galaxies), spiral arms ($4$ galaxies), plausible tidal signatures ($7$ galaxies), and gravitational strong-lensing arc ($1$ galaxy). We also note that our galaxies span two observational depth regimes -- CANDELS/Wide ({\it HST} $2$-orbit; $6$ galaxies) and CANDELS/Deep ({\it HST} $10$-orbit; $10$ galaxies) depths {as described in \cite{grogin_candels:_2011,Koekemoer}}. For each galaxy in our sample, we extract residual substructures using the images with best-available depth, unless otherwise specified. In Figure\,\ref{clean_pTF_sTF}, we provide example postage-stamps of our test set galaxies and their respective residual features to be extracted and analyzed in this paper. For reference, we also show an example `clean' (i.e., visually no left-over) residual image.

	\begin{table*}
		\centering
		\caption{Test sample of galaxies hosting different residual substructures which we use to demonstrate our feature extraction technique. Columns: (1) the CANDELS field identifier -- UDS, GOODS-South (GDS), GOODS-North (GDN), EGS, followed by the galaxy's official CANDELS identification number; (2--3) celestial coordinates in degrees; (4) the best-available redshift for the galaxy \citep[see][for details]{dahlen_critical_2013}; (5) the stellar mass \citep{santini_stellar_2015,Mobasher15}; (6) the type of residual substructure hosted by the galaxy; (7) the observational depth based on the official CANDELS exposure maps as described in {\protect \cite{Koekemoer}}. $^\dagger$ denotes that galaxy is involved in a close-pair system. $\ddagger$ indicates that we use the galaxy to demonstrate a deep vs. shallow feature-extraction comparison.}
		\label{table_can_test_sample}
		\begin{tabular}{ccccccc}
			\hline
			ID & R.A. & Decl. & $z_{\rm best}$ & $\logten(M_{\rm stellar}/M_{\odot})$ & Residual Feature & Observational Depth\\
			(1)  & (2) & (3) & (4) & (5) & (6) & (7)\\
			\hline
			UDS 4165 & $34.580786$ & $-5.25419$ & $0.985$ & $10.63$ & spiral &  shallow \\
			UDS 24437$^\dagger$ & $34.339863$ & $-5.149623$ & $1.823$ & $11.2$ & plausible tidal& shallow \\
			GDS 4608$^\dagger$ & $53.081307$ & $-27.871586$ & $1.067$ & $10.84$ & plausible tidal & deep$^\ddagger$ \\
			GDS 14637 & $53.04408$ & $-27.785048$ & $1.045$ & $10.11$ & plausible tidal & deep$^\ddagger$ \\
			GDS 14876 & $53.11879$ & $-27.782818$ & $2.309$ & $10.91$ & plausible tidal & deep$^\ddagger$ \\
			GDS 16440$^\dagger$ & $53.050929$ & $-27.772409$ & $1.033$ & $10.63$ & plausible tidal &  deep\\
			GDS 18160 & $53.163589$ & $-27.758952$ & $1.095$ & $10.82$ & spiral & deep \\
			GDS 18959 & $53.202349$ & $-27.751279$ & $1.113$ & $10.91$ & disk & deep \\
			GDN 7220 & $189.172373$ & $62.191548$ & $1.019$ & $11.03$ & spiral & deep\\
			GDN 14758$^\dagger$ & $189.358539$ & $62.241522$ & $1.021$ & $9.8$ & plausible tidal & deep \\
			GDN 19863 & $189.154657$ & $62.274604$ & $2.482$ & $10.55$ & plausible tidal & deep\\
			EGS 1690 & $215.092284$ & $52.922091$ & $1.081$ & $10.7$ & spiral & shallow \\
			UDS 10713 & $34.275997$ & $-5.22157$ & $0.627$ & $11.11$ & lensing arc & shallow \\
			EGS 10518 & $214.877124$ & $52.819527$ & $1.192$ & $10.72$ & disk & shallow \\
			EGS 12782 & $215.20786$ & $53.064729$ & $1.103$ & $10.19$ & disk & deep\\
			EGS 27018 & $215.006664$ & $52.996171$ & $1.244$ & $10.51$ & disk & shallow\\
			\hline
		\end{tabular}
	\end{table*}
	
	\vspace*{0.5cm}
	\subsection{VELA Datasets }
	\label{vela_data}
	To demonstrate our new methodology on the extraction of residual substructure signatures hosted by a galaxy-galaxy major merger, we apply it to mock images from VELA\#01\footnote{The VELA simulation-suite comprises of 35 intermediate-mass galaxies ($\log_{10}(M_{\rm stellar}/M_{\odot}) \sim 10$ at $z\sim 2$) and they are referred using an ID starting from $01$ to $35$.} of the VELA zoom-in hydro-dynamic simulations \citep{Ceverino14,zolotov15}. 
	Briefly, in this simulation baryonic physics\footnote{The VELA simulations include many physical prescriptions related to galaxy evolution -- gas cooling, star formation, ISM metal enrichment, photo-ionization heating from the Ultra-Violet background radiation, stellar-mass loss, stellar feedback, tracing of smooth cosmological accretion of other galaxies and gas.} are solved in conjunction with gravitational N-body interactions using the Adaptive Refinement Tree (ART) code \citep{kravtsov97,kravtsov03,Ceverino09} with spatial resolution of $17.5\,{\rm pc}$ to $35\,{\rm pc}$. The VELA simulation-suite has been extensively used in the context of galaxy morphology such as elongated galaxies \citep{Ceverino15b,Tomassetti16}, giant clumps \citep{Moody14,Inoue16,Mandelker17}, compaction \citep{zolotov15,Tacchella16,Huertas-Company18}, galaxy spin and size \citep{Jiang18}, H-alpha morphology \citep{Ceverino16a,Ceverino16b}, and structural parameters \citep{snyder15}. Starting at a cosmological scale factor $a= 0.125$ ($z\sim7$), the simulation information is recorded at equal increments of $\Delta a = 0.01$, which approximately corresponds to a time frame $\Delta t=100\,{\rm Myr}$ at $z=2$. For the merging system VELA\,$01$, we use the information during $15$ time intervals spanning $0.37\leq a \leq0.47$ ($1.7\lesssim z \leq 1.1$), bracketing roughly $0.6$ ($0.8$) Gyr before (after) the coalescence stage (occurring at $a=0.41$; $z=1.44$). We note that this system experiences a gas-rich, major merger (stellar mass ratio $\sim1.33:1$, gas mass ratio $\sim 1:1$) resulting in post-merger stellar and gas masses of $\logten{(M_{\rm stellar}/M_{\odot})} = 10.05$ and $\logten{(M_{\rm gas}/M_{\odot})}=9.55$, respectively.
	
	At each time step, we use mock {\it HST} WFC3/F160W observations of the merging system generated using the methodology {described in \citet{snyder15,Simons19}}. Briefly, they use the raw-simulation information at each time-step and assign stellar population synthesis model-informed \citep[][IMF]{BC03,Chabrier03} spectral energy distributions (SEDs) to the stellar particles using {\tt SUNRISE} \citep{jonsson06,jonsson10} with dust scattering and absorption assuming a dust-to-metals ratio of $0.4$ \citep{Dwek98} and a Milky Way-like dust grain size distribution \citep[$R_{\rm v} = 3.1$;][]{Weingartner01,Draine07}. At each time-step, they integrate this 3-dimensional data onto a 2D camera plane to produce an idealized image at  19 (camera) orientation angles \citep[for details see Table\,1 in][]{Huertas-Company18}, which are then convolved with the point spread function (PSF) generated by {\tt Tiny Tim} \citep[][WFC3/F160W]{Krist11} to reach desired spatial resolution. {These mock idealized images are available on MAST\footnote{\url{https://archive.stsci.edu/prepds/vela/}} \citep[][DOI: \url{https://doi.org/10.17909/t9-ge0b-jm58}]{Simons19}}.

	To match the empirical CANDELS F160W $5\sigma$ limiting surface-brightness sensitivities at both Wide ($25.25\,{\rm mag\,{arcsec^{-2}}}$) and Deep ($26.25\,{\rm mag\,{arcsec^{-2}}}$) depths, we add Gaussian random noise to the F160W PSF convolved images of VELA\,$01$. This process of generating mock {\it HST} CANDELS images from simulated data products is often referred to as {\it CANDELization} \citep{Huertas-Company18} and we call the resultant depth-matched images as {\it CANDELized} images.  We apply {\tt GALFIT} to the VELA\,$01$ {\it CANDELized} images to generate single-S\'ersic residual images consistent with the vdW12 empirical data at each time step, viewing angle, and depth realization for a total of $380$ unique residual images.

	\begin{figure*}
		\includegraphics[width=2\columnwidth]{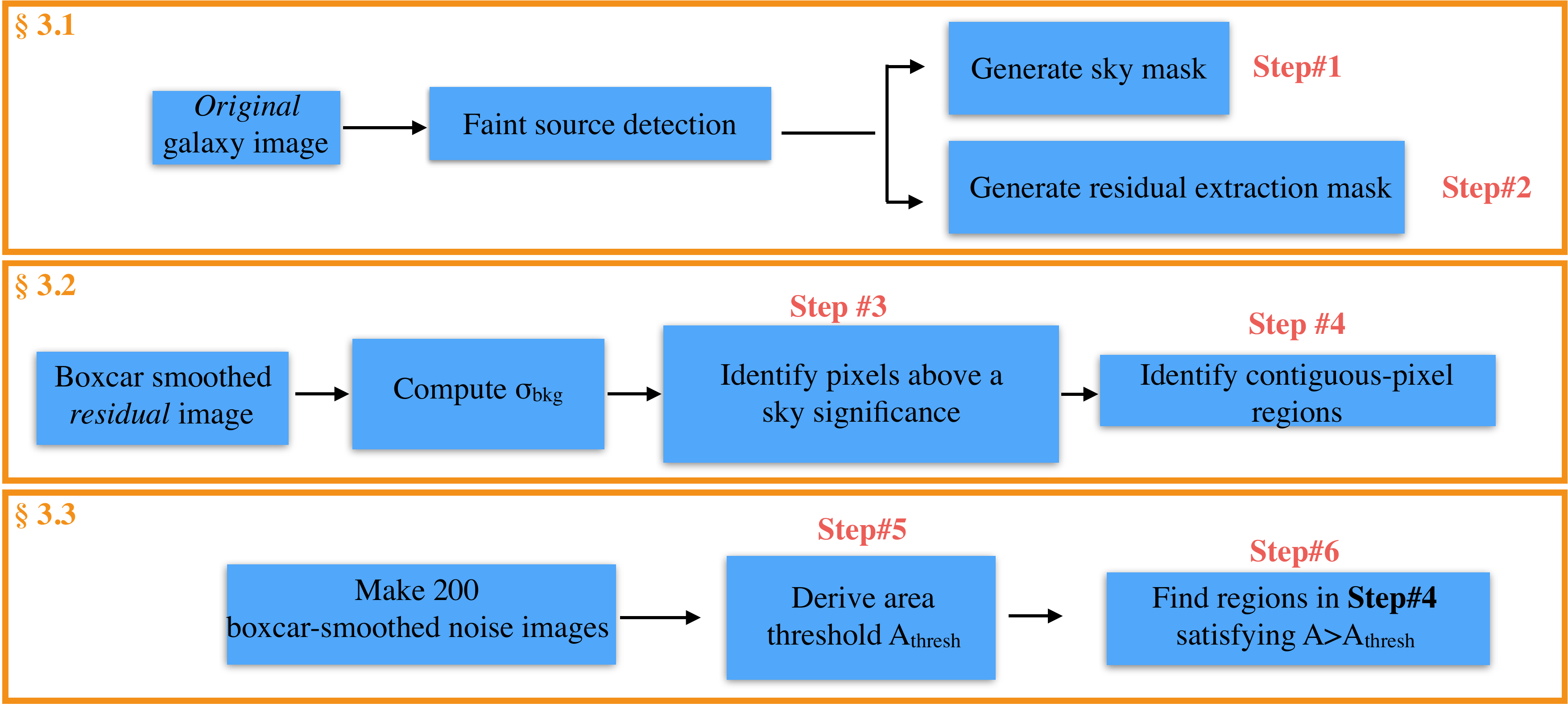}
		\caption{Flow diagram illustrating our detailed step-by-step residual substructure extraction process, which is divided into three main parts (orange boxes), described in their respective sub-sections (\cref{methodology}). We refer to the key steps (red text) when presenting our method in \cref{methodology}. }
		\label{flow_chart}
	\end{figure*}
	
	
	\begin{figure*}
		\includegraphics[width=2\columnwidth]{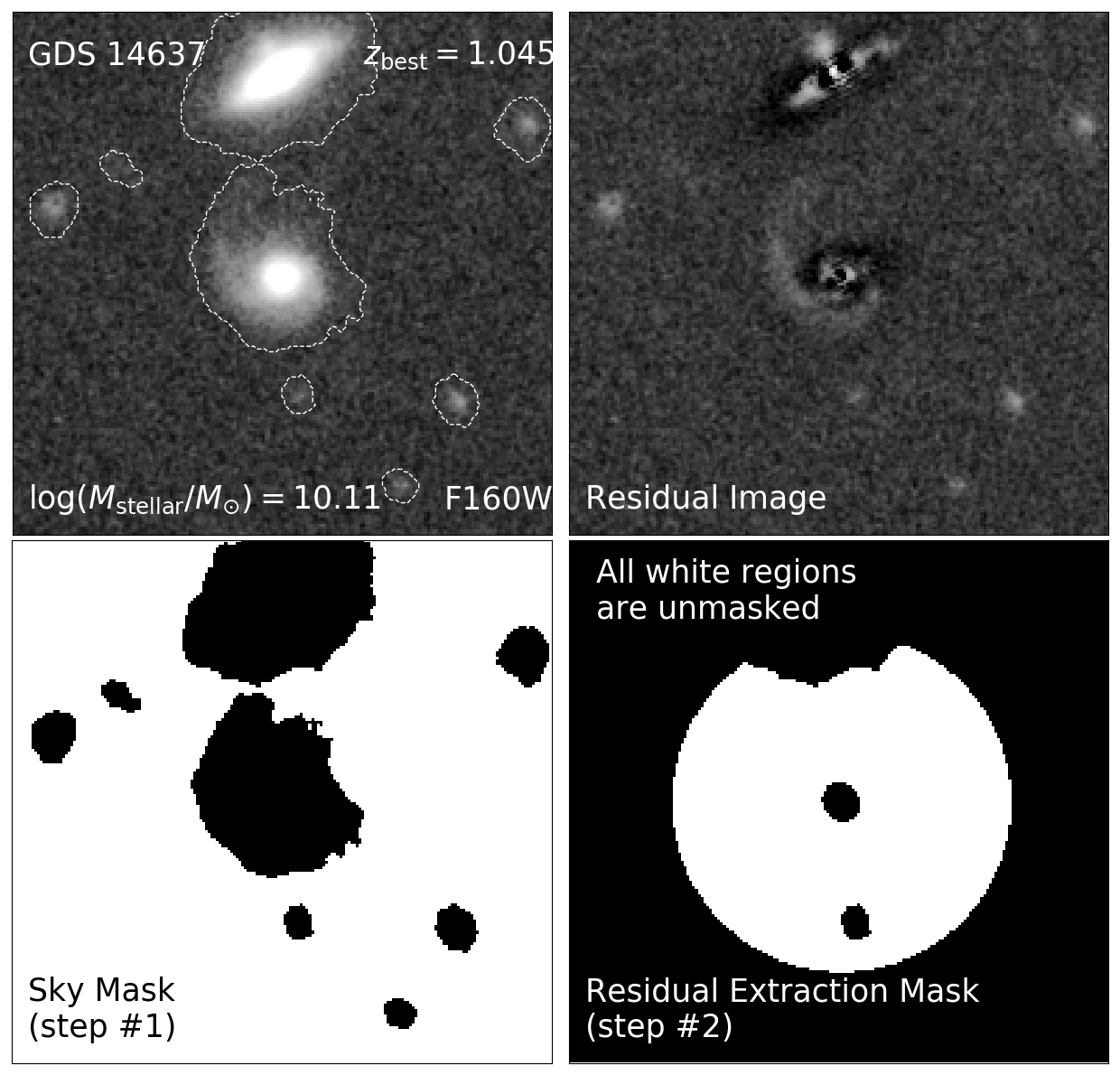}
		\caption{Visualization of the dual masking scheme described in \cref{masking_and_sky} for an example galaxy (GDS $14637$; at the center of image; see Table\,\ref{table_can_test_sample}).  {\it Top left:} WFC3/F160W ($1.6\,\mu{\rm m}$) image with faint source detections shown with white outlines. {\it Top right:} single-S\'ersic based residual image from vdW12. {\it Bottom left:} mask generated by step\#1 for measuring background sky in unmasked regions (white; pixel values$=1$); masked regions are shown in black (pixel value$=0$). {\it Bottom right:} mask generated by step\#2 for extracting residual substructure. }
		\label{source_detect_and_masks}
	\end{figure*}
	
	\section{Methodology}
	\label{methodology}
	We demonstrate our residual substructure extraction method by analyzing the {\tt GALFIT}-based residual images of a selective sample of galaxies hosting different left-over substructures (\cref{candels_data}). 
	We organize the functionality of our method into three main objectives as follows. Starting with a {\it H}-band original image, we generate two masks to facilitate estimation of local-sky background ($\sigma_{\rm bkg}$) and residual feature extraction. Next, we estimate $\sigma_{\rm bkg}$ in the smoothed single-S\'ersic-based residual image and search (in proximity to the galaxy-of-interest) for contiguous-pixel regions that satisfy a sky-significance. Finally, we identify the flux-wise significant regions whose areas are larger than a pixel-contiguity area threshold ($A_{\rm thresh}$) informed using noise-only Monte-Carlo (MC) simulations to quantify their {\it H}-band surface brightness. We present a detailed flow-chart of our residual extraction process in Figure\,\ref{flow_chart} and label the key steps to be discussed in this section.
	
	{We acknowledge that our methodology is closely dependent on the outputs generated by different software routines such as GALFIT and Source Extractor. Although these software are extensively used in the astronomy community, their specific functionality may change in the future updated versions. In this context, we emphasize that our modular algorithmic approach described in \cref{methodology} and illustrated in the flow chart (Figure\,\ref{flow_chart}) can be treated as a road map for future applications using the updated or new versions of the software analogous to GALFIT or Source Extractor.}
	\begin{table}
		\centering
		\caption{Source extraction configuration parameters to inform our masking routine. Columns: (1) name of the parameter; (2) value employed in this study.}
		\label{table_source_identification}
		\begin{tabular}{cc}
			\hline
			Parameter & Value Used\\
			(1) & (2)\\
			\hline
			DETECT\_THRESH & 0.7\\
			MIN\_AREA & 7 \\
			DETECT\_NTHRESH & 32\\
			DEBLEND\_MINCONT & 0.0001\\
			CONV\_FILTER & 2D Tophat (radius = $5$)\\
			\hline
		\end{tabular}
	\end{table}

	\subsection{Masking} 
	\label{masking_and_sky}
	We devise a dual masking scheme to facilitate the extraction and isolation of residual substructures at radii outside the each galaxy center. This requires robust estimation of the local-sky background, detection of faint sources not fit by vdW12, and a definition of galaxy center. The magnitude criterion imposed by vdW12 to mask nearby faint objects (\cref{candels_data}) is a reasonable choice in the context of quantifying galaxy structural properties. However, such faint un-fit sources can have comparable surface brightness to our desired faint residual substructure (\eg\, tidal features).
	In addition, central galactic regions may house complex substructures (especially in young post-mergers) that may not be well represented by a S\'ersic profile and may lead to significant central left-over light \citep[\eg\,][]{Hopkins09} that can obscure the detection of faint tidal features of interest. 
	
	\subsubsection{ {\bf (Step\#1)} Generate sky mask}
	\label{source_identification}
	We start by performing faint source detection within the postage stamp of the galaxy 
	(original image) using a python-based implementation \cite[{SEP};][]{barbary16} of the standard source-extraction tool SExtractor \citep{bertin96}. 
	We find that the source extraction parameter values provided in Table\,\ref{table_source_identification} yield optimal identification of small faint sources in the images, without deblending faint extended features (\eg\, plausible tidal signatures) into separate objects\footnote{A small value of {DETECT\_THRESH} ensures that the resultant segmentation maps envelop the outer-most light of the sources An alternative to this step is to use a higher detection threshold value and grow the resultant segmentation regions to fully encompass the extended light.}.
	Compared to the SExtractor configurations used to generate the CANDELS source catalogs \citep[\eg\, see][]{galametz_candels_2013,guo_candels_2013}, we employ a smaller convolution filter size (five vs. nine) for better detection of smaller and fainter sources. 
	
	In the top panels of Figure\,\ref{source_detect_and_masks}, we demonstrate the thorough detection of all faint sources seen in the vdW12 residual image of an example galaxy (GDS $14637$; center of image). Hereafter, we use this galaxy to demonstrate all of our analysis steps outlined in the flow-chart (Figure\,\ref{flow_chart}). In the bottom left panel, we provide the resultant background sky mask. Masked regions are shown in black (pixel value$=0$). The unmasked regions (white) represent the sky contribution in the vicinity of the galaxy, which use during our feature extraction in \cref{extracting_features}.
	
	

	\subsubsection{{\bf (Step\#2)} Generate residual extraction mask}
	We also generate a mask for extraction of residual substructures within a physically-motivated proximity of $R_{\rm proj} < 30\,{\rm kpc}$ from the centroid of the galaxy-of-interest.
	This corresponds to nearly five (ten) times the effective size of late (early) type galaxies at $z\sim 1$ \citep{van_der_wel_3d-hst+candels:_2014}. 
	We find that this extraction radius encompasses all faint residual substructure for the galaxies in our sample. As shown in Figure\,\ref{source_detect_and_masks} (bottom-right panel), we mask contributions
	from the central core regions of each galaxy residual so that we extract only outer features. We define the inner masking by an ellipse centered on the galaxy, with semi-major and semi-minor axes equal to 1.5 times those of the SExtractor detection ellipse from step\#1, and orientation along the SExtractor position angle. Finally, we also mask any faint companions from step\#1 that fall within the $R_{\rm proj} < 30\,{\rm kpc}$ extraction region. In summary, 
	we use this mask during the extraction of residual substructures (\cref{extracting_features}).
	
	
	\begin{figure*}
		\centering
		\includegraphics[width=2\columnwidth]{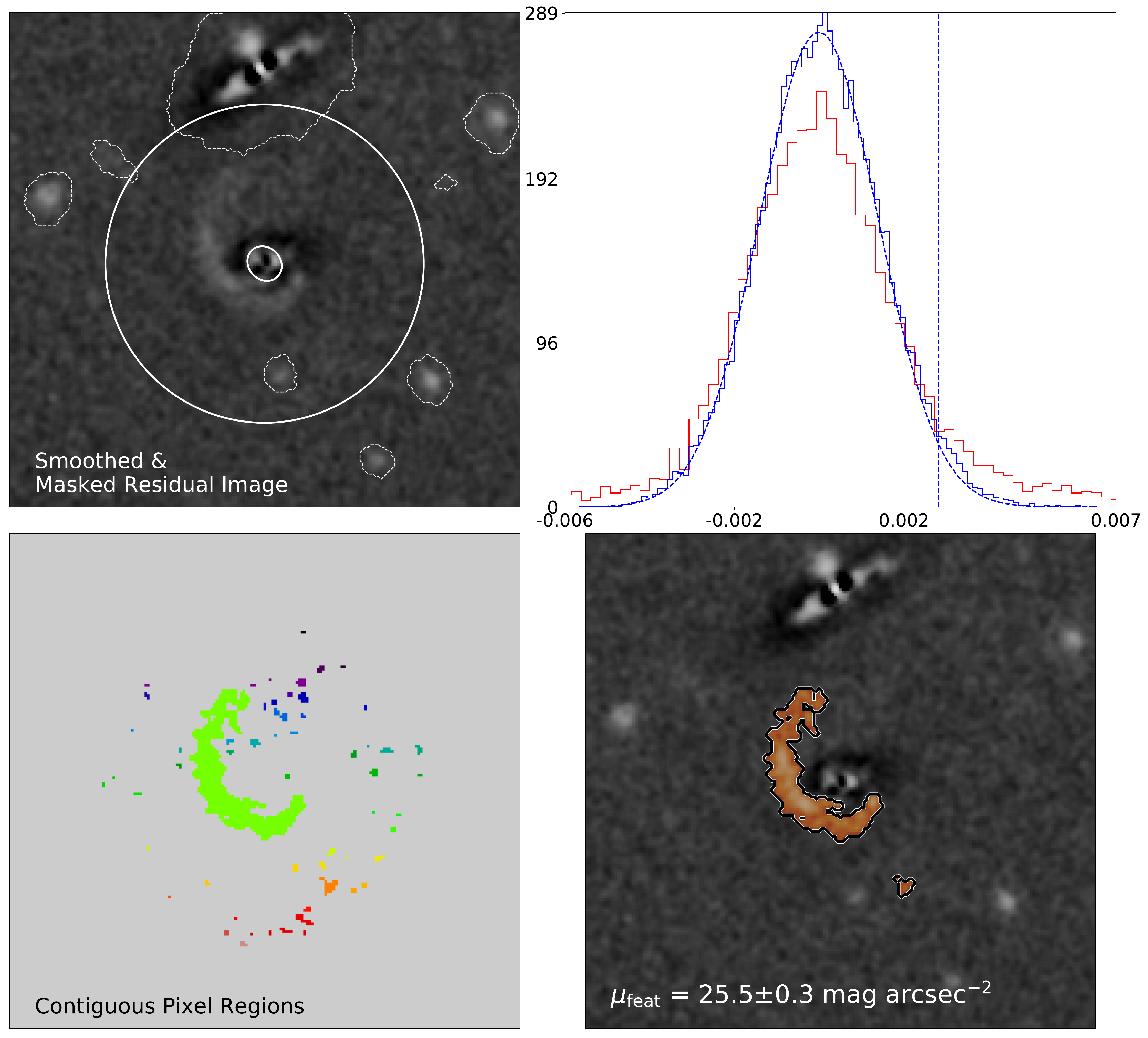}
		\caption{Visualization of the residual feature extraction steps described in \cref{extracting_features,MC_area_contribution} for an example galaxy (shown in Figure\,\ref{source_detect_and_masks}). {\it Top-left:}  2D Boxcar kernel-smoothed ($3\,{\rm pix}\times3\,{\rm pix}$; side length $\sim 0.18''$) residual image with the inner and outer residual feature mask region boundaries (step\#1) given in solid white, and faint-source masking (step\#2) shown as dashed white outlines. {\it Top-right:} the normalized pixel-value distribution of the sky background (blue histogram) in units of electrons/sec, its corresponding best-fit Gaussian curve (dashed-blue curve), and $2\sigma_{\rm bkg}$ value (vertical blue line) are compared to the distribution of pixel values from the residual extraction region (red histogram). {\it Bottom-left:}  the pixels residing in the residual extraction region satisfying a flux-wise significance cut $f_{\rm pix}>2\sigma_{\rm bkg}$ (step\#3) are color coded according to their unique contiguity (described in detail in \cref{identify_contiguity_pixels}; step\#4). {\it Bottom-right} the final residual features (red shading) after rejecting smaller area regions as described in \cref{MC_area_contribution} (steps\#5 and \#6); the feature surface brightness and its 68\% confidence level are given.}
		\label{fig_can_overview_of_process}
	\end{figure*}
	
	\subsection{Final Extraction of Residual Features}
	\label{extracting_features}
	We start the feature extraction process by analyzing a Boxcar-smoothed version of the residual image ({\it top-left panel}; Figure\,\ref{fig_can_overview_of_process}) to facilitate the robust identification of faint features. 
	We utilize the masks described in \cref{masking_and_sky} to identify all unmasked pixels within the residual extraction mask region that are
	significantly above the background sky. Then, we identify contiguous groupings of such pixels to define unique residual features.
	
	\subsubsection{{\bf (Step\#3)} Identify flux-wise significant pixels}
	\label{flux_wise_sig_pix}
	The first step in extracting residual features is to quantify the background sky. We compute the background $\sigma_{\rm bkg}$ as the standard deviation of the best-fit Gaussian profile representing the normalized distribution of pixel values in the unmasked regions of the smoothed residual image after applying the sky mask (\cref{masking_and_sky}). In Figure\,\ref{fig_can_overview_of_process} ({\it top-right panel}), we show the sky pixel-value distribution (blue histogram) with its corresponding Gaussian curve. We denote the $2\sigma_{\rm bkg}$ with a vertical dashed line.
	
	Next, we apply the residual extraction mask and identify all unmasked pixels in the smoothed-residual image (red histogram in  Figure\,\ref{fig_can_overview_of_process}, {\it top-right panel}). In Figure\,\ref{fig_can_overview_of_process} ({\it bottom-left panel}), we plot all such pixels satisfying a fiducial flux-wise significance cut $f_{\rm pix}>2\sigma_{\rm bkg}$, where $f_{\rm pix}$ is flux value of the pixel. Visually, we observe one large flux-wise significant residual feature and many small areas of $>2\sigma_{\rm bkg}$ pixels. We note that a less stringent significance cut of $f_{\rm pix}>1\sigma_{\rm bkg}$ increases both the number of small pixel groupings and the typical area per pixel-group, albeit with decrease in the overall signal-to-noise ratio. On the other hand, a more strict $f_{\rm pix}>3\sigma_{\rm bkg}$ cut identifies fewer pixels (with high signal-to-noise), but with significant structural discontinuities (\ie\, an extended feature is broken into multiple smaller regions). As such, we elect to use $f_{\rm pix}>2\sigma_{\rm bkg}$ as our fiducial criterion for flux-wise significant pixels in our analysis.
	
	
	\subsubsection{{\bf (Step\#4)} Identify contiguous-pixel regions}
	\label{identify_contiguity_pixels}
	To isolate and separate larger, contiguous residual substructures from smaller groupings of flux-wise significant pixels, we use a python-based 2D pixel-connectivity algorithm {\tt skimage: label} \citep{Fiorio96,Wu05}.  This tool assigns unique identification number to each contiguous pixel region using a binary image in which all the flux-wise significant pixels are set to unity and the remainder are set to zero. In Figure\,\ref{fig_can_overview_of_process} ({\it bottom-left panel}), we show the contiguous flux-wise significant pixels (color coded regions).
	
	Next, we compute the area and enclosed flux of these contiguous regions using a companion module {\tt skimage: regionprops}\footnote{The module computes cumulative flux per unique contiguous region in the residual image at their respective pixel coordinates.}  \citep{Reiss93,Burger09}. Re-affirming our visual interpretation, we find a contiguous substructure with large area (shown in green) in the immediate vicinity of our example galaxy and many regions with smaller areas (Figure\,\ref{fig_can_overview_of_process}; {\it bottom-left panel}). We note that correlated-noise fluctuations can cause several pixels to be flux-wise significant and contiguous by random chance. We quantify this effect and reject such spurious regions in the following section.

	\subsection{Final Feature Selection}
	\label{MC_area_contribution}
	We isolate the final, flux-wise and area-wise significant residual features following a MonteCarlo approach. We estimate an area threshold ($A_{\rm thresh}$) based on a noise-only pixel contiguity expectation and select those flux-wise significant contiguous features with areas larger than $A_{\rm thresh}$.
	
	\subsubsection{{\bf (Step\#5)} Quantify area threshold $A_{\rm thresh}$}
\begin{figure}
		\includegraphics[width=1\columnwidth]{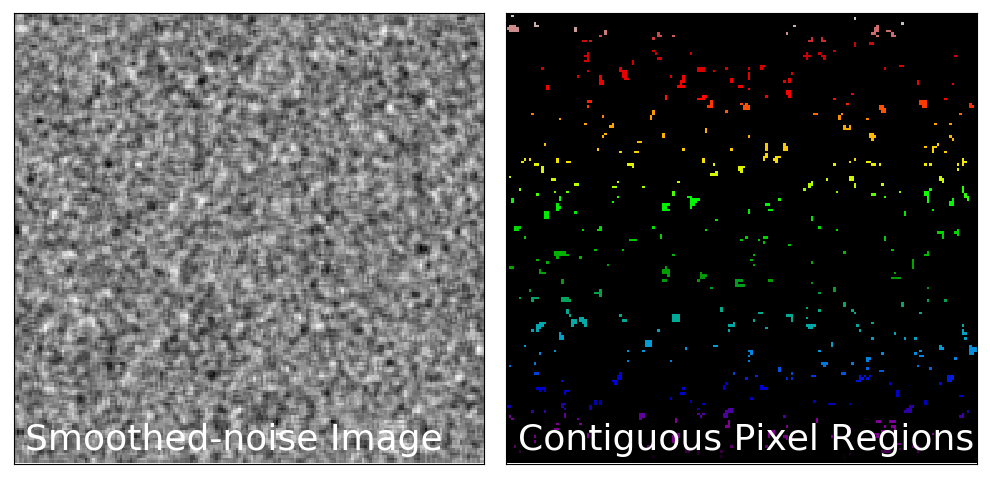}
		\caption{Visualization of the pixel contiguity by random chance (step\#5). {\it Left panel:} the Boxcar smoothed random noise image for one randomly chosen MonteCarlo iteration generated as described in \cref{MC_area_contribution}. {\it Right panel:} the contiguous pixel regions satisfying our fiducial flux-wise significance cut $f_{\rm pix}>2\sigma_{\rm bkg}$ (color coding).}
		\label{visualize_mc_sims}
	\end{figure}

	As a first step to quantify an area threshold above which the noise contribution to pixel contiguity is minimum, we generate $200$ Boxcar smoothed images with Gaussian noise matching the $\sigma_{\rm bkg}$ value from step\#3. In each random-noise image, we identify contiguous regions satisfying our fiducial flux-wise significance and quantify their area (as described in \cref{flux_wise_sig_pix}).  In Figure\,\ref{visualize_mc_sims}, we show an example smoothed random noise image ({\it left panel}) and its corresponding flux-wise significant contiguous regions ({\it right panel}; color coding). We notice that the noise-generated contiguous areas visually mimic the smaller area regions identified in step\#4 (Figure\,\ref{fig_can_overview_of_process}; {\it bottom-left panel}). Combining the result of 200 MonteCarlo iterations for our example galaxy, we find an extraction area threshold $A_{\rm thresh} = 20$, where $> 99\%$ of the contiguous pixel regions in smoothed noise-only images have areas smaller than $20$ pixels. We repeat this MonteCarlo process for all galaxies in our analysis that span different CANDELS fields and find a range of $A_{\rm thresh}$ values between $17$ to $23$ pixels, owing to varying field-to-field noise properties.

	\subsubsection{{\bf (Step\#6)} Identify and Quantify Final Residual Features}
	To select final significant features, we identify the contiguous regions with areas larger than the threshold set by the noise properties (\ie\, $A>A_{\rm thresh}$). We then quantify the surface brightness of the final feature regions ($\mu_{\rm feat}$) as:
	\begin{equation}
	\mu_{\rm feat} = -2.5\logten(F) + zp + 2.5\logten(A^{\prime})~[{\rm mag\,arcsec^{-2}}]~,
	\end{equation}
	where $F$ is the cumulative flux (in Jansky) of all the regions satisfying our criteria, $zp = 8.9$ is the zero-point magnitude, and $A^{\prime}$ is the cumulative area of all these significant contiguous regions in units of arcsec$^{2}$. We also report the photometric error on the quantified surface brightness by adding the noise from $\sigma_{\rm bkg}$ and Poisson noise from the feature in quadrature. In Figure\,\ref{fig_can_overview_of_process} ({\it bottom-right panel}; shading), we show the final residual features (an extended tidal-tail structure) and quantify its $H$-band surface brightness to be $\mu_{\rm feat} = 25.5\pm0.3\,{\rm mag\,arcsec^{-2}}$.
	
	\begin{figure*}
		\centering
		\includegraphics[width=2.1\columnwidth]{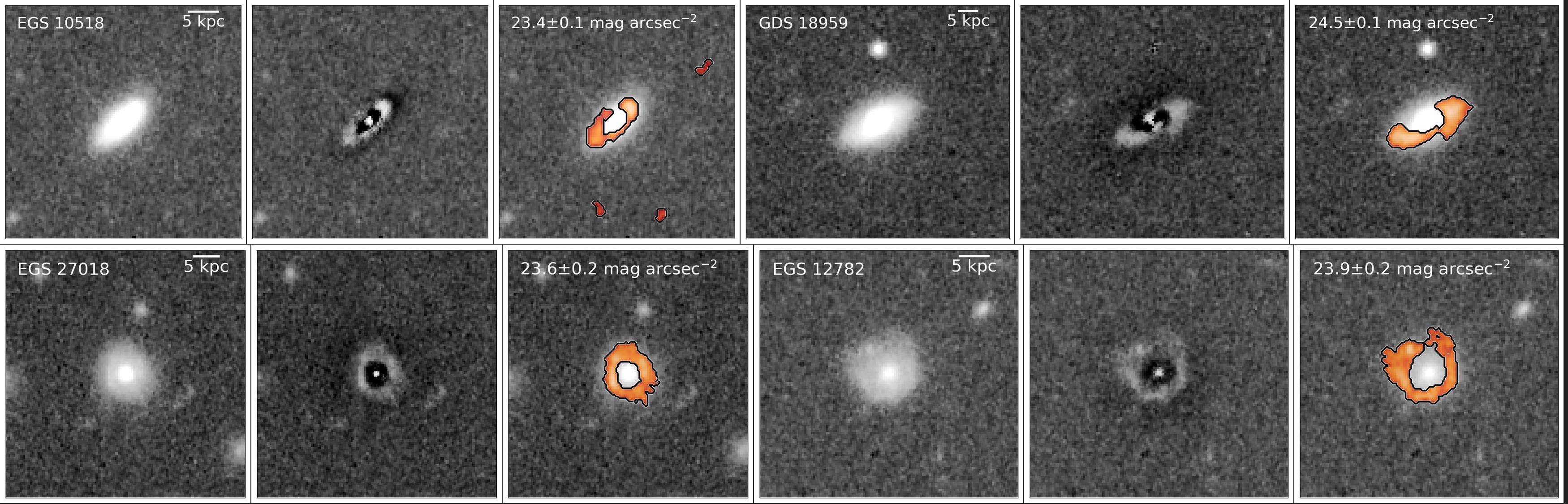}\\
		\includegraphics[width=2.1\columnwidth]{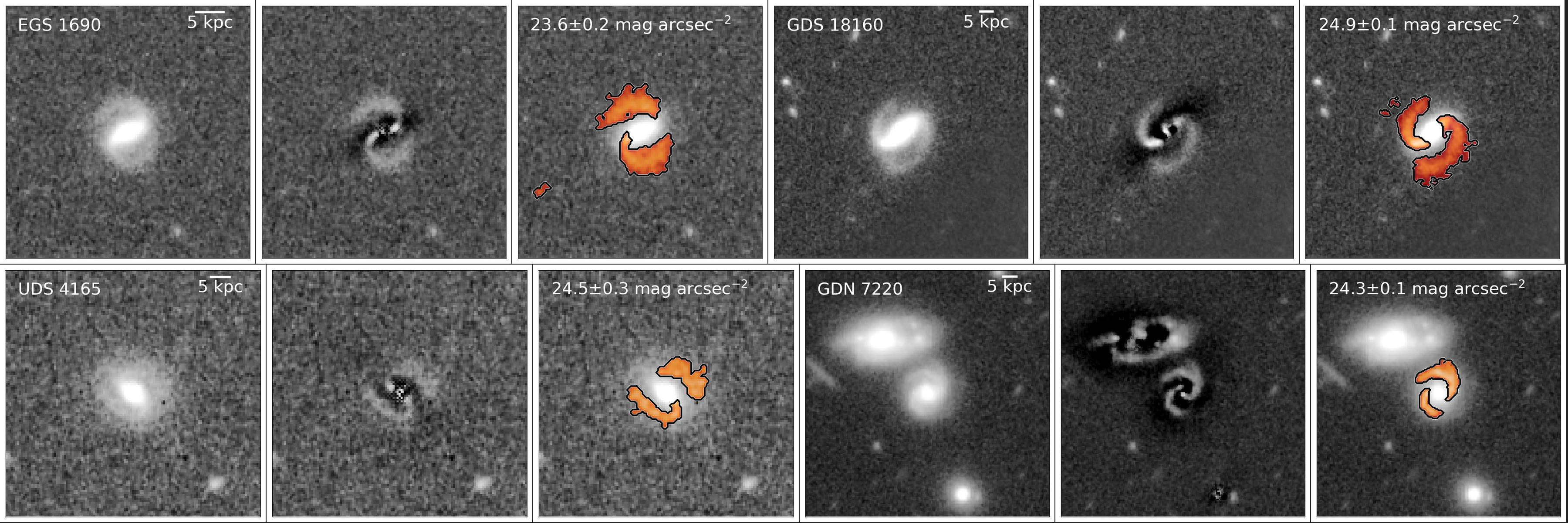}
		\caption{Example residual feature extraction and quantification for select CANDELS galaxies exhibiting disk structure ({\it top-two rows}) and spiral substructure ({\it bottom-two rows}). For each galaxy, we show the {\it original} WFC3/F160W ($1.6\,\mu{\rm m}$) image (left) and include the official CANDELS ID and a $5\,{\rm kpc}$ physical scale; the single-S\'ersic residual image from vdW12 (center); and the extracted residual features (solid-black outlines with red shading) overlaid on its respective {\it original} image (right), and we include the feature surface brightness measurement.}
		\label{collage_disk_spiral}
	\end{figure*}

	\section{Application to CANDELS Galaxies}
	\label{extract_res_struc_can}
	In this section, we demonstrate the application of our residual substructure extraction method on example CANDELS galaxies hosting a range of features -- disk, spiral, and plausible interaction signatures (see Table\,\ref{table_can_test_sample}). We also illustrate the role of image depth by comparing results at {\it HST} $2$-orbit and $10$-orbit depths for three galaxies (GDS\,$4608$, GDS\,$14637$, and GDS\,$14876$).

	\subsection{Extracting disk and spiral features}
	\label{spiral_disk_features}
	In Figure\,\ref{collage_disk_spiral} ({\it top-two rows}), we show the results of applying our feature extraction method on galaxies visually identified to be hosting residual disk structure. In the cases of  EGS\,$10518$ and GDS\,$18959$, we find similar looking, inclined disk-like residual structures, whereas the residuals of EGS\,$27018$ and EGS\,$12782$ exhibit circular features (likely owing to their face-on orientation). We also note hints of clump-like substructures embedded within the disk feature of EGS\,$12782$. We find that these features have surface brightness measures ranging between $ 23.4 - 24.5\,{\rm mag\,arcsec^{-2}}$.
	Re-affirming our visual interpretation, we find that our method extracts features that are consistent with a residual disk, likely owing to an underlying bulge$+$disk morphology being fit with a single-S\'ersic model.
	
	In Figure\,\ref{collage_disk_spiral} ({\it bottom-two rows}), we show the results of our feature extraction on galaxies hosting residual spiral substructures (EGS\,$1690$, GDS\,$18160$, UDS\,$4165$, and GDN\,$7220$). We extract two-arm spiral patterns for all the four galaxies and find their surface brightness ranges between $23.6-24.9\,{\rm mag\,arcsec^{-2}}$. We note that the spiral substructures exhibited by GDS\,$18160$ and GDN\,$7220$ are visually apparent (despite having fainter surface brightness) than EGS\,$1690$ and UDS\,$4165$, owing to the difference in their observational depths (deep vs. shallow; see Table\,\ref{table_can_test_sample}). We also note that all except GDN\,$7220$ appear to be hosting a central bar. Extracting such spiral substructures with our method will help future analyses quantify their physical parameters such as pitch angle, bar structure.

	\begin{figure*}
		\centering
		\includegraphics[width=2.1\columnwidth]{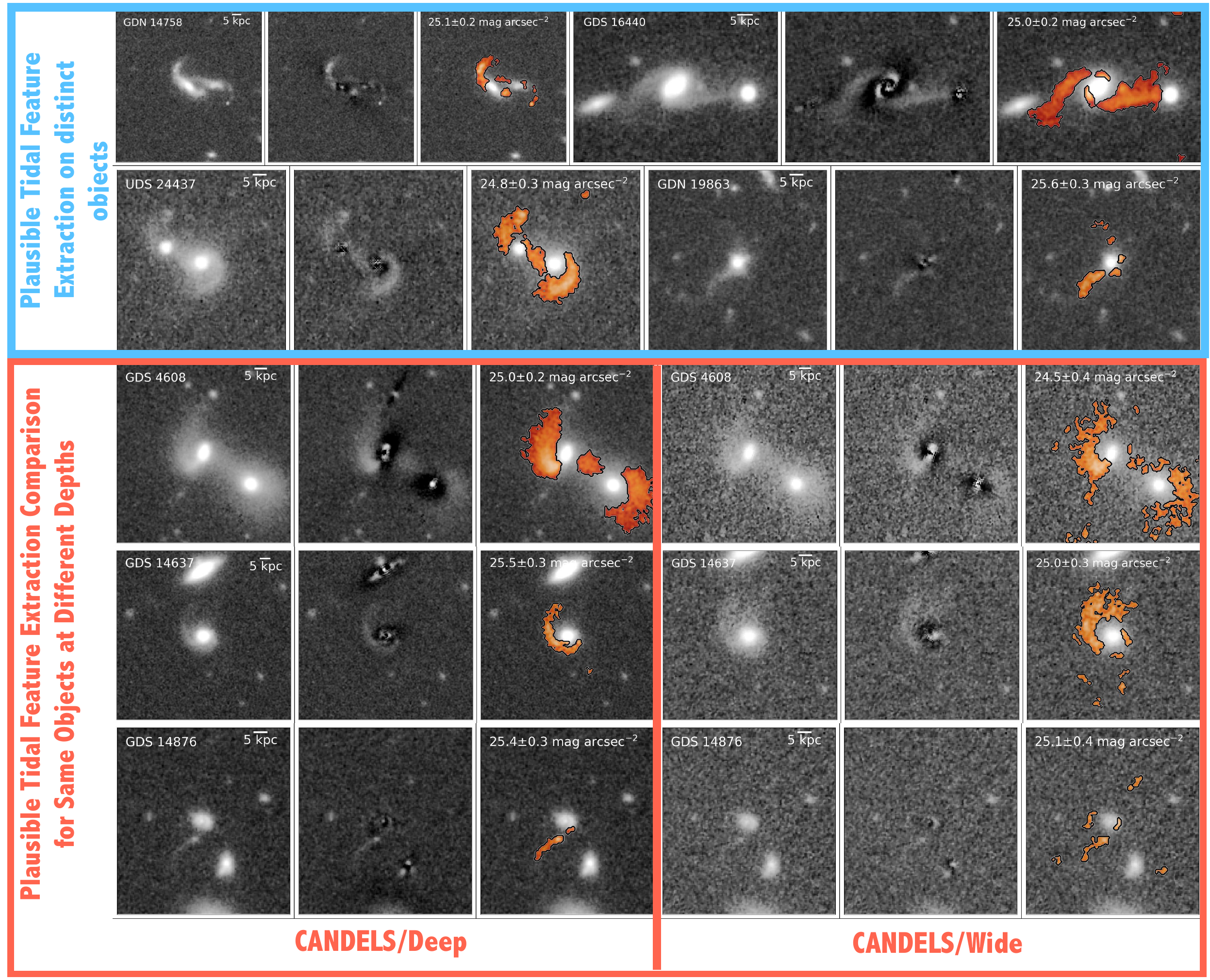}\\
		\caption{ Example residual structure extraction and quantification for select CANDELS galaxies hosting plausible tidal signatures. For each galaxy, we show the {\it original} image, single-S\'ersic residual image, and the extracted residual features using the same format as in Figure\,\ref{collage_disk_spiral}. For three example galaxies (bottom three rows), we compare the feature extraction process at CANDELS Deep ({\it HST} 10-orbit; left panels) and Wide ({\it HST} 2-orbit; right panels) depth configuration.}
		\label{collage_ptf}
	\end{figure*}

	\subsection{Extracting plausible tidal features}\label{extracting_plausible_tf}
	In Figure\,\ref{collage_ptf}, we show the  results of our feature extraction method on galaxies exhibiting plausible tidal interaction signatures. These galaxies broadly fall into one of the following four categories -- i) close-pair systems in which both galaxies exhibit plausible signatures of tidal interaction (GDN\,$14758$, UDS\,$24437$, and GDS\,$4608$); ii) close-pair systems in which only one galaxy exhibits a plausible interaction feature (GDS\,$16440$ and GDS\,$14637$ ); one example galaxy experiencing a plausible minor interaction (GDS\,$14786$);  and one plausible post-merger system (GDN\,$19863$). The extracted residual substructures in these examples highlight our visual interpretations.
	
	First, we apply our feature extraction method to each galaxy in the three example close-pair systems with plausible dual interaction signatures. We note that all three pairs have $|\Delta z|/(1+z) \leq 0.025$ and residual substructure consistent with theoretical predictions for tidal features produced by major mergers \citep{Duc2013}. We find a striking structural similarity between the features exhibited by UDS\,$24437$ -- UDS\,$23110$ pair ($z\sim 1.8$) and the GDS\,$4608$ -- GDS\,$4529$ pair ($z\sim 1$), such that each close pair hosts two tidal arms and an overlapping region between the host and companion galaxy. These broad and extended features are similar to those found in observations and simulations of major mergers between spheroid-dominated galaxies \citep[\eg\,][]{bell_dry_2006}. The residual substructure midway between each galaxy core in these two pairs may be a stellar bridge or may be an artifact of the simultaneous S\'ersic fitting of two galaxies in very close proximity. In contrast, each galaxy in pair GDN\,$14758$--GDN\,$14516$ appears to have two tidal arms which are thin and clumpy in some places. The characteristics of the latter tidal arms are likely the result of interacting disk-dominated galaxies. 
	
	Next, we demonstrate the feature extraction on two pairs in which only one galaxy exhibits plausible interaction signatures. GDS\,$14637$ has a single arm-like residual that is similar in appearance to the broad and extended tidal arms discussed above. In contrast, GDS\,$16440$ shows two arm-like features, where one extends outward away from the host and is aligned with a nearby (compact) galaxy (at right) with a spectroscopic redshift difference of only $|\Delta cz|\sim 100\,{\rm km\,s^{-1}}$ from GDS\,$16440$. This picture is consistent with theoretical merger simulations \citep[e.g., Figure\,6 from][]{Duc2013} of a disk-dominated galaxy experiencing a major merger with a nearby compact companion. We note that GDS\,16440 has a clear inner spiral pattern that extends to a radius of $R\sim5$--10\,kpc. Despite its point-like appearance suggesting a plausibly interaction-fueld AGN, the companion has no known X-ray detection to corroborate this suggestion. Finally, we acknowledge the possibility that the extracted arm-lke features of GDS\,$16440$ may be extensions of the inner spiral structure, yet, their diffuse appearance and physical extent out to $R>25$\,kpc is much more consistent with tidal interaction signatures. Admittedly, quantitatively distinguishing spiral versus interaction features is critical for robust identification of galaxy merging. While beyond the scope of this work, we believe that further quantification of extracted residual features will provide improved methods for distinguishing distinct physical processes.
	
	Finally, we demonstrate our residual feature extraction technique on two additional galaxies with plausible signs of tidal activity. First, GDS\,$14786$ shows a minor companion galaxy with a narrow $\sim 10\,{\rm kpc}$ tail-like feature, consistent with numerical predictions of features caused by minor mergers  \citep{Namboodiri85,Duc2013}.
	Second, GDN\,$19863$ appears to have two tidal-tail residuals, each extending $\sim 10\,{\rm kpc}$ from the galaxy center. One is much fainter than the other. Multiple faint tidal tails are found briefly after coalescence in simulations of major mergers.
	
	\subsection{Role of Depth during Feature Extraction}
	\label{role_of_depth}
	
	To illustrate the role of imaging depth on the extraction and quantification of {\it different} plausible tidal features, we compare the results of residual substructures extracted from deep (CANDELS/Deep) and shallow (CANDELS/Wide) images for three examples in Figure\,\ref{collage_ptf}.  We perform this comparison on the following examples from \cref{extracting_plausible_tf}: GDS\,$4608$ (close-pair system where both galaxies show plausible tidal arms), GDS\,$14637$ (close-pair system where on galaxy shows plausible interaction feature), and GDS\,$14876$ (galaxy experiencing a plausible minor interaction).
	
	We find that the morphologies of the extracted residuals at a shallow $2$-orbit depth are qualitatively similar to their $10$-orbit depth counterparts in all the three cases, albeit with noticeably noisy visual-appearance. We note that the features extracted from the shallow depth images tend to be confined to the brighter surface brightness (by $0.3-0.5\,{\rm mag\,arcsec^{-2}}$) regions of the same features extracted from the deep images. The shallow residuals are also lower in signal-to-noise when compared to features extracted at a deeper depth. This is expected as an increase in noise per pixel reduces the structural contiguity, owing to the reduction in number of pixels meeting flux-wise significance cut, thereby, resulting in a less significant yet brighter surface brightness.  Although we re-calibrate the feature extraction employing a larger area threshold value for shallow depth images, we notice additional small faint sources in the nearby projected vicinity that are contaminating the extracted features. This is because such faint sources are not significantly detected during the source extraction and masking steps, but are extracted from the residuals by our method. This deep-shallow comparison highlights the role of  observational depth in identifying faint, high-redshift plausible tidal signatures.

	
	\section{Extracting Tidal Features in a major merger from cosmological simulation}
	\label{extract_tf_vela}
	In this section, we demonstrate the application of our residual substructure extraction method on synthetic {\it HST} images of a major merger from the VELA simulation \citep{Ceverino14}. We extract the residual features and quantify their areas during $10$ time steps of the merging process spanning $1.7\lesssim z \leq 1.1$ ($\Delta t\sim1.4\,{\rm Gyr}$), each at $19$ viewing orientation angles, and two observational depth configurations matching empirical CANDELS observations (Wide and Deep). We note that the first pericentric passage and the coalescence stages occur at $z=1.63$ and $z=1.44$, respectively. To visually comprehend the quantitative trends, we also discuss the qualitative appearance of example extracted features at one fixed viewing orientation.
	
	\subsection{Quantitative Analysis of Extracted Features}
	\label{quant_interp_vela}
\begin{figure}
		\centering
		\includegraphics[width=1\columnwidth]{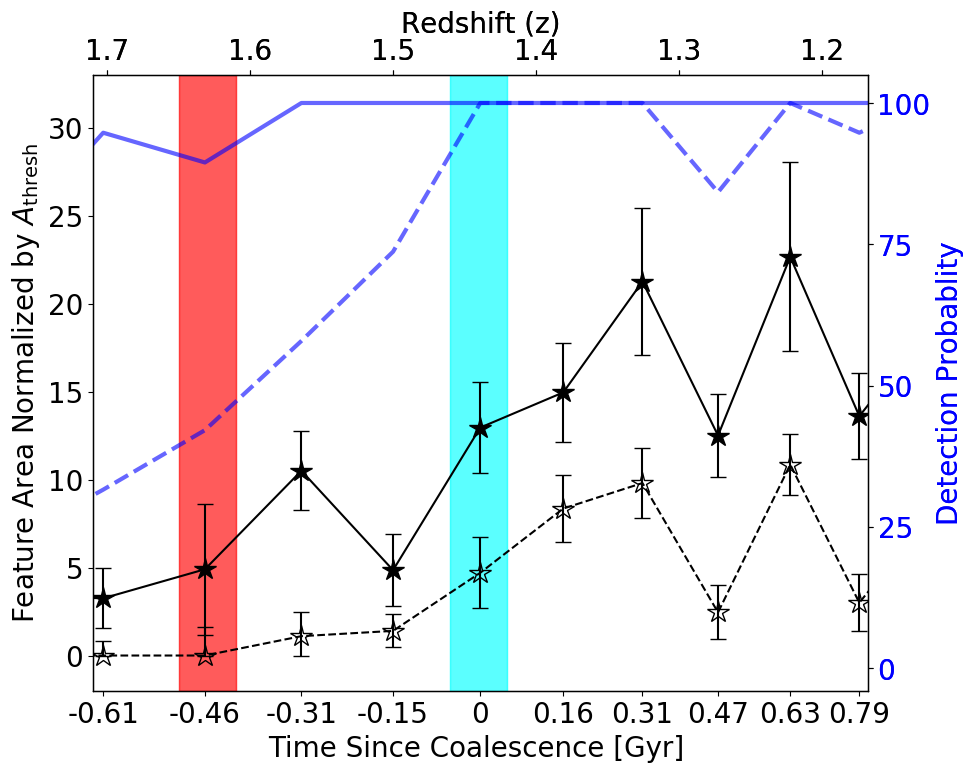}
		\caption{Area of the extracted residual features with $A\geq A_{\rm thresh}$ and $\mu_{\rm feat}$ brighter than a depth-dependent limiting surface brightness, as a function of time since the coalescence phase from the synthetic mock observations of a galaxy merger simulation. We normalize the feature area with $A_{\rm thresh}$ value employed for each CANDELS depth -- Wide (filled points) and Deep (open points). At a fixed time step and depth, we show the median feature normalized area with a star; its corresponding standard deviation across different viewing orientations is shown with a error-bar.  In blue, we give the detection probability which is defined as the percentage of camera angles during which the extracted residual features satisfy our area and surface brightness criteria at shallow depth (dashed line) and deep depth (solid line). We emphasize the first pericentric passage and coalescence stages with red and cyan shading, respectively.}
		\label{vela_sb_v_time}
	\end{figure}

	In Figure\,{\ref{vela_sb_v_time}}, we show the median extracted feature areas as a function of time before and after the coalescence stage ($t=0\,{\rm Gyr}$) for CANDELS Wide and Deep depths at all $19$ viewing angles. We also quantify the detection probability as the percentage of viewing sightliness during which we detect features with their areas larger than the $A_{\rm thresh}$ and $\mu_{\rm feat}$ brighter than the limiting surface brightness (Wide -- $25.25\,{\rm mag\,arcsec^{-2}}$;  Deep -- $26.25\,{\rm mag\,arcsec^{-2}}$). We find the deep and shallow extracted residuals follow qualitatively similar trends in terms of their changing {\it prominence} (i.e., feature area) during different pre- and post-merger stages. Owing to coarse time resolution ($\sim 150\,{\rm Myr}$) between each time step, we are only able to discuss broad trends and differences between the deep and shallow results. 
	
	At deeper image depth, we detect residual substructure at nearly all viewing angles and time steps. Prior to first pericentric passage ($t=-0.46\,{\rm Gyr}$), the feature area is minimal ($A = 3\pm 1 \,A_{\rm thresh}$). At first pericentric passage, we find a plausible increase in the feature prominence, albeit with a large spread between different viewing orientations. This increase clearly peaks at the next time step 150\,Myr later, hinting at the formation of tidal interaction-triggered features. If true, these features abruptly diminish in significance (by a factor of two) at $t=-0.15\,{\rm Gyr}$, after which extracted feature areas steadily increase through the coalescence stage until they peak at 300\,Myr later with a value 3--7 times greater than the initial minimum. This temporal behavior strongly implies the presences of tidal merger signatures. During the three time steps that follow ($0.47\leq t \leq 0.79$), we note that the feature prominence follows a stochastic rising and falling trend with a range of area values between those of coalescence and the last maximum, or even slightly more at some viewing angles. As we discuss later, these stochastic fluctuations during different time steps correspond to noticeable changes in the visual appearance of the extracted residual features.

	In the case of shallow depth images, we typically find a similar temporal time evolution trend, albeit with some key distinctions. First, we note that the trends at shallow depths are shifted to systematically smaller areas when compared to the same features observed in deeper images. This is because of the decreased signal-to-noise per pixel causing less number of contiguous pixels to satisfy our feature extraction criteria. Next, contrary to what we observe in deeper images, we find no increase in the prominence of the extracted feature at the $t=-0.31\,{\rm Gyr}$ time step following first pericentric passage. This is consistent with the idea that tidal interaction features are faint and their detection is highly sensitive to image depth. Indeed, we find the detection probability of any residual substructure in the shallow images is less than 50\% until after first passage, and only reaches unity at the coalescence stage when the first noticeable increase in feature area occurs. During $t>0\,{\rm Gyr}$, we note that post-merger time steps with detection probability $<100\%$ correspond to smaller feature areas. Likewise, maximum feature areas correspond to high detection probability. These subtle trends highlight the role played by viewing lines-of-sight during the detection of certain merger features, especially in shallow depth images. Nevertheless, during the time steps with high detection probabilities (irrespective of the depth), we note that the viewing angles can induce a scatter of $20-40\%$ in the measured feature prominences.

	\begin{figure*}
		\includegraphics[width=2.1\columnwidth]{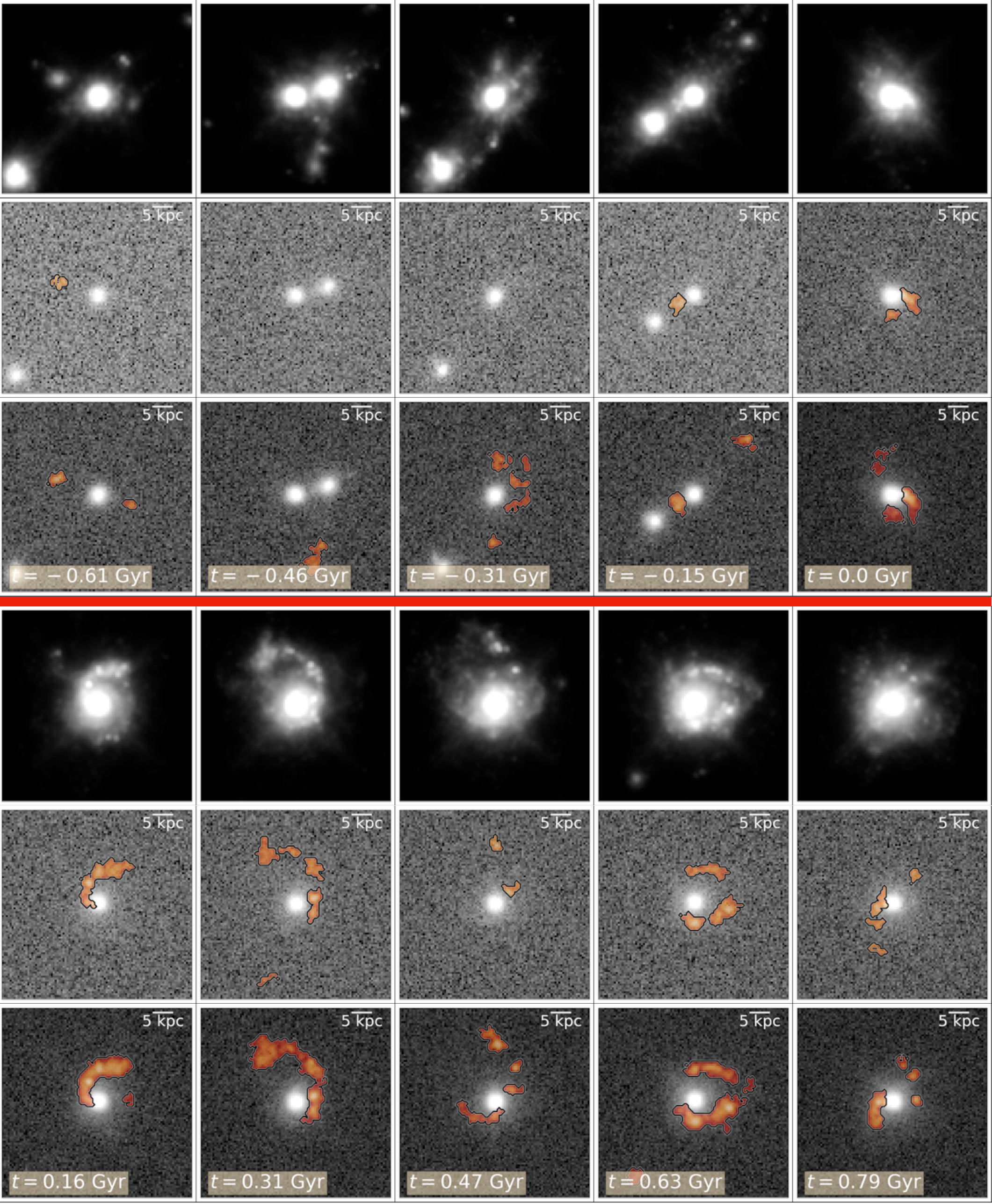}\\
		\caption{ Extracted residual features during the 10 time steps shown in Figure\,\ref{vela_sb_v_time} from a galaxy merger simulation (VELA; see text for details). At each time step (left to right), we present the idealized (noise-less) mock {\it HST} WFC3/F160W images for one viewing angle (cam$00$; {\it top}), the CANDELS/Wide depth images ({\it middle}), and the CANDELS/Deep depth images ({\it bottom}).
			The extracted features are shown with red shading for the shallow and deep data.
			For each trio of images, we indicate the $5\,{\rm kpc}$ physical scale and the time elapsed since (or prior to) the coalescence stage ($z=1.44$; $t=0\,{\rm Gyr}$).}
		\label{collage_vela}
	\end{figure*}

	\subsection{Qualitative Interpretation of Extracted Features}
	\label{qual_interp_vela}
	To illustrate the temporal quantitative trends shown in Figure\,\ref{vela_sb_v_time}, we discuss the qualitative appearances of the features extracted from one example viewing orientation. We show the extracted features at shallow and deep depths alongside their idealized images during $10$ time steps in Figure\,\ref{collage_vela}. We discuss the extracted features from different image depths with respect to the distinct structures evident in their idealized (no noise added) images, and we compare the morphologies of the features extracted from the VELA mock data to those from CANDELS images of plausible merging systems shown in Figure\,\ref{collage_ptf}.
	
	During the first two time steps ($t\leq -0.46\,{\rm Gyr}$), we extract small residual structures in both deep and shallow images that correspond to faint sources in proximity to the merging galaxies within their idealized images. At $t=-0.31\,{\rm Gyr}$ ($\sim 150\,{\rm Myr}$ after the first pericentric passage), we extract features at deeper depths that correspond to the interaction debris surrounding the merging galaxies  in the idealized images. However, we do not detect these features in shallow depth images, which aligns with the noticeable difference in feature prominence at $t=-0.31\,{\rm Gyr}$ in Figure\,\ref{vela_sb_v_time}. Just prior to coalescence ($t=-0.15\,{\rm Gyr}$), we extract a small feature midway between the two merging galaxies at both depths, which corresponds to a plausible bridge-like structure in the idealized image. We find similar excess-light features in the extracted residuals for UDS\,$24437$ and GDS\,4608 in Figure\,\ref{collage_ptf}. Yet, interestingly, the VELA merger does not exhibit tidal arms during its close-pair stage, which is in contrast to all three example plausible interacting systems (GDS\,$4608$, UDS\,$24437$, and GDN\,$14758$).

	Starting at coalescence ($t=0\,{\rm Gyr}$), we extract features at both depths that highlight the birth of a post-merger tidal-arm that grows in area over the next $\sim 300\,{\rm Myr}$. This feature is similar in appearance to the single arm-like structure we extract from one of the two example CANDELS pairs in which only one galaxy has a plausible interaction signature (GDS\,$14637$; Fig.\,\ref{collage_ptf}). As such, GDS\,14637 is also consistent with a post-merger in close proximity to another galaxy.
	At $t=0.47\,{\rm Gyr}$, we extract two narrow tail features with clumpy visual appearance in the deep image, which resemble the plausible post-merger signature we extract for GDN\,$19863$. This picture is consistent with the theoretical expectations that major gas-rich mergers produce diffuse tidal features with clumpy knots \citep{Fensch17}. We note that we only marginally detect a fraction of this tidal feature in the shallow depth image, which aligns with the abrupt decrease in feature prominence noted in Figure\,\ref{vela_sb_v_time}. Finally, during the last two time steps in both shallow and deep images we extract a large and extended plausible tidal loop or two-tailed feature (at $t=0.63\,{\rm Gyr}$) that becomes less prominent at $t=0.79\,{\rm Gyr}$. The changes in these features during the last two time steps illustrate the stochastic fluctuations in the extracted feature areas we find for all viewing angles in Figure\,\ref{vela_sb_v_time}. This exercise illustrates the interplay between interaction/merger feature extraction and observational depth.

	\section{Discussion}
	\label{discussion}
	So far, we have presented the development of a residual substructure extraction tool and discussed its application to example galaxies from the CANDELS survey and mock {\it CANDELized} observations of a galaxy merger simulation. In this section, we discuss several additional possible applications of our residual extraction method.

	\subsection{Possible Applications}
	We present some possible applications of our residual feature extraction method towards improving popular structural-fitting routines, studying multi-wavelength properties of different single-component S\'ersic residual features (\eg\, disk structures from host bulge$+$disk systems) and substructures (\eg\, spiral arms, plausible interaction signatures), and extracting strong gravitational arcs.
	
	\subsubsection{Improving structural fitting} 
	An important step in many galaxy structural-fitting studies \citep[\eg][]{Bruce12,van_der_wel_3d-hst+candels:_2014} is to choose the number of required parametric components during {\tt GALFIT} modeling (or other similar tools), starting with a simple single-S\'ersic model and advancing to more complex models (\eg\, bulge$+$disk) including even the inclusion of Fourier modes (\eg\, to model spiral features). For upcoming large-scale surveys, developing machinery to make such decisions is key for automated structural modeling for very large samples. One way to approach this is to use machine learning to inform the number of parametric components required to best represent a galaxy morphology \citep[\eg][]{Tuccillo18}. Alternatively, we propose that our method can be used to inform additional residual components and repeat  {\tt GALFIT} with updated number of components. 
	
	\subsubsection{Multi-wavelength application} 
	Empirical studies often perform multi-wavelength decomposition of galaxy structure into their bulge and disk components to separately inform their physical parameters \citep[\eg\, stellar mass, color;][]{Bruce14,Dimauro18}, which can provide key insights into the physics of galaxy evolution. We propose that extending our feature extraction method to multiple wavelengths can inform the physical properties for disk, central bulge, spiral arms, and interaction signatures. Furthermore, recent studies are developing resolution-element-based spectral energy distribution fitting in conjunction with spectroscopic measurements to produce maps of physical parameters \citep[stellar-mass, star formation rates;][de la Vega et al. in prep, Jafariyazani et al. in prep]{Wuyts13}. Using our method in conjunction with these semi-resolved maps can enable future investigations to study structurally-resolved properties (e.g., colors, star-formation, chemical composition, kinematics) of different galactic substructures \citep[especially for tidal features; \eg\, see][]{Kado-Fong18}.

	\subsubsection{Extracting Gravitational Arcs}
	\label{extract_grav_arcs}
	Yet another possible application of our method is the extraction of strong gravitational lensing arcs. Strong gravitational lensing provides an unique opportunity to observe the otherwise inaccessible high-redshift universe. Several previous studies have used strong lensing to study cosmology, the physical properties of lensing systems, and the properties of distant background lensed sources \citep[\eg\,][]{Refsdal64,Kochanek95,Treu04}. Identifying and extracting the gravitational lensing signatures is a key step in such works, which often use {\tt GALFIT} to subtract the host galaxy light to unearth such lensing features \citep[\eg\,][]{Brammer12,van_der_Wel13}.
	
\begin{figure*}
		\centering
		\includegraphics[width=2\columnwidth]{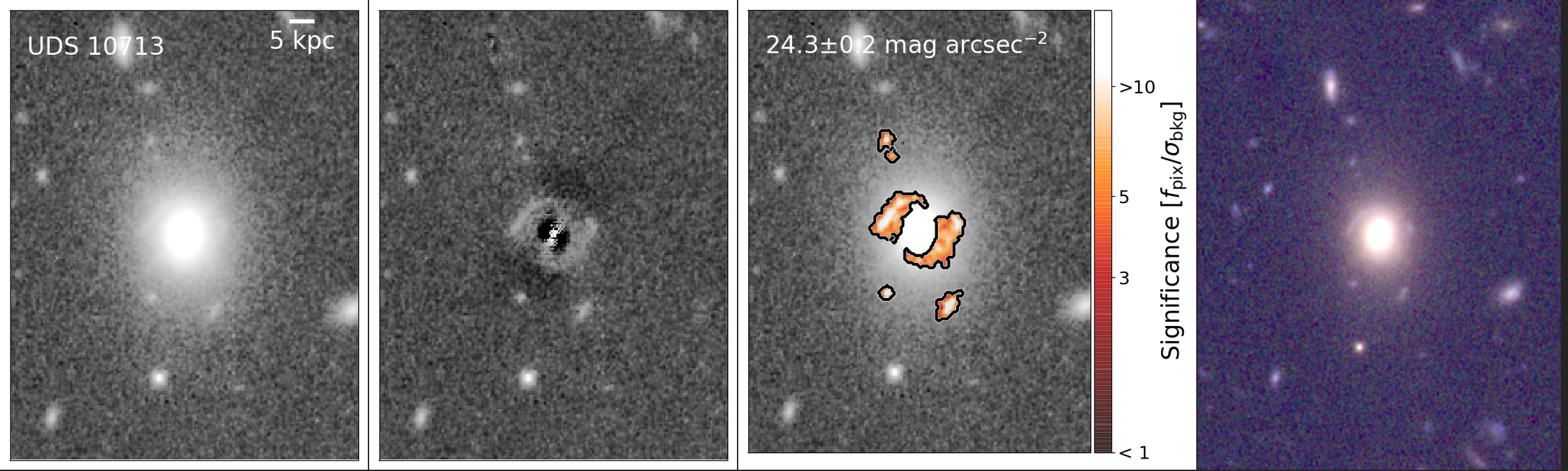}
		\caption{Example extraction of a candidate strong lensing arc hosted by the CANDELS galaxy UDS\,$10713$ \citep[center of the image; identified by][]{Hocking18}. {\it First panel:} the WFC3/F160W ($1.6\mu{\rm m}$) {\it original} image with the official CANDELS ID and a $5$ kpc physical scale. {\it Second panel:} the single-S\'ersic residual image by vdW12. {\it Third panel:} the extracted residual features (shading) overlaid on the {\it original} image; the signal-to-noise is indicated with the color-bar. {\it Fourth panel:} false-color F160W--F125W--F606W RGB image of  UDS\,$10713$; strong lensing arc appears at the top-left of UDS\,$10713$.}
		\label{gravitational_arcs}
	\end{figure*}

	In Figure\,\ref{gravitational_arcs}, we show the result of our feature extraction method on an example CANDELS strong-lensing candidate (UDS\,$10713$), which is identified using a machine learning based classification scheme by \cite{Hocking18}. We use the galaxy's single-S\'ersic residual image by vdW12 and extract significant contiguous substructure above a $3\sigma_{\rm bkg}$ threshold (instead of our fiducial $2\sigma_{\rm bkg}$). We find arc-like features with surface brightness $\mu_{\rm feat}=24.3\pm0.2\,{\rm mag\,arcsec^{-2}}$ in close vicinity to the galaxy, which qualitatively aligns with the blue-arc in its corresponding false-color RGB image (at roughly $10$ o'clock). However, we note that the single-S\'ersic residual of UDS\,$10713$ shows excess galactic light and additional sources in its projected proximity, which are contributing to the surface brightness of our extracted features. Isolating such contaminants is beyond the scope of this current work and we recommend that future studies should employ more complex S\'ersic models for better galactic light subtraction and use multi-wavelength information of the features for better automated lensing feature detection and extraction. Furthermore, future works can extend our residual extraction method to galaxy cluster environments where strong lensing can be more prevalent (usually with high magnifications) as a means to derive cluster mass measurements.
	
	\subsection{Future Prospects with the James Webb Space Telescope}
	\label{future}
\begin{figure}
		\includegraphics[width=1\columnwidth]{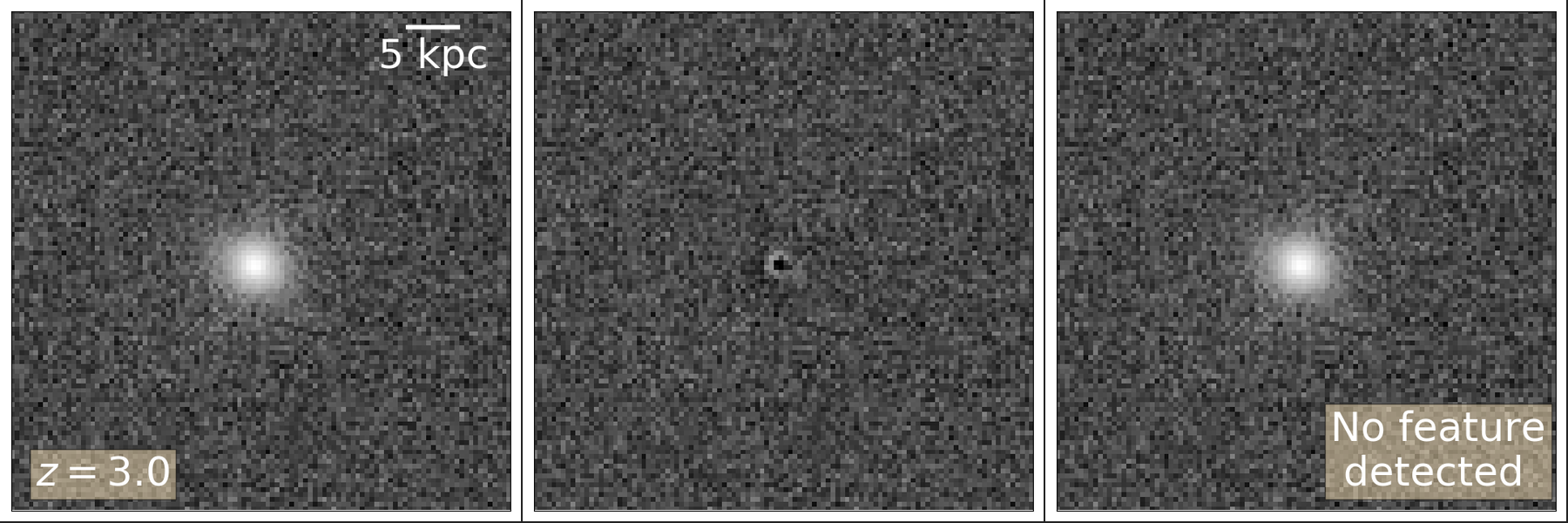}\\
		\hspace{0.05cm}\includegraphics[width=1\columnwidth]{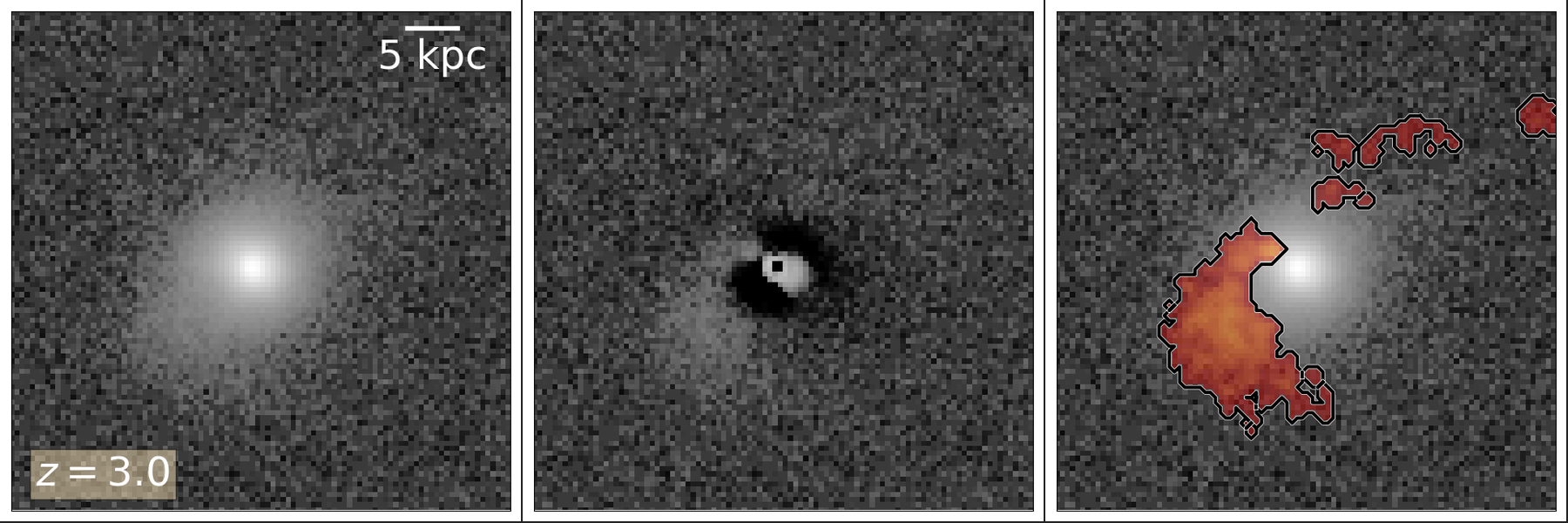}
		\caption{ A simulated massive post-merger at $z=3$ (VELA $21$; see text for details) with mock CANDELS/Deep {\it HST} imaging ({\it top row}), and mock {\it JWST}-ERS CEERS imaging ({\it bottom row}). In each row, we present the synthetic noise-added images matching their respective empirical depths ({\it left}), the corresponding single-S\'ersic GALFIT residual images ({\it middle}), and the extracted feature (if detected) shown with red shading ({\it right}). This figure demonstrates that {\it JWST} will enable us to probe tidal features out to high redshifts in a considerably smaller amount of exposure time and at longer rest-frame wavelengths when compared to {\it HST}.}
		\label{can_vs_jwst}
	\end{figure}

	In preparation to probe high-redshift galaxy merging using the James Webb Space Telescope ({\it JWST}), we analyze synthetic {\it JWST} observations of a different merger in the VELA cosmological simulation suite (VELA\,$21$), which experiences a major merger (post-merger stellar mass $\log_{10}(M_{\rm stellar}/M_{\odot}) =10.5$; stellar-mass ratio $\sim 3:1$) at $z\sim 3.2$ (coalescence). In Figure\,\ref{can_vs_jwst}, we compare and contrast the feature extraction process on its mock {\it JWST} and {\it HST} observations. 
	
	Motivated by our simulated observations of a major merger in \cref{extract_tf_vela},  we focus this exercise on one specific merger stage $z=3$ ($\sim120\,{\rm Myr}$ after coalescence), where we expect the galaxy to exhibit tidal features. We generate idealized mock {\it HST} WFC3/F160W and {\it JWST} NIRCam/F277W images using the process described in \cref{vela_data}. We add Gaussian noise to them such that they match the CANDELS/Deep depth and the proposed Cosmic Evolution Early Release Science Survey (CEERS; PI: S. Finkelstein, {\it JWST} PID: 1345) depth \footnote{The proposed depth of the CEERS survey reaches greater than a $10\sigma$ point-source and $5\sigma$ extended-source sensitivity of $28\,{\rm mag}$ with a $\sim 2800$ second exposure in NIRCam/F277W.}, respectively. Finally, we generate the single-S\'ersic residuals by applying {\tt GALFIT} to the {\it HST} and {\it JWST} noise-added images following vdW12, and apply our feature extraction technique on these residuals (\cref{methodology}).
	
	While we detect no significant residual features at CANDELS/Deep {\it HST} depth, we extract extended tidal-fan features in the NIRCam/F277W {\it JWST} CEERS depth image, which mimic the structures exhibited by our example CANDELS close-pair system (GDS\,$4608$; see Figure\,\ref{collage_ptf}). Our demonstration asserts the capabilities of {\it JWST} to probe faint tidal features out to high-redshifts with considerably less exposure time and at a longer rest-frame wavelengths than the current extremes accessible to {\it HST}. This will enable future studies to probe high-redshift merging systems that are key towards understanding the role of galaxy merging in galaxy evolution.

	\section{Conclusions}
	\label{conclusions}
	As a {\it first step} towards quantitatively identifying `hallmark' signatures of galaxy merging (\eg\, tidal arms, tails, bridges, and extended fans), in this paper, we introduce a new analysis tool to extract and quantify residual substructures hosted by galaxies. Our method analyzes residual images produced by any standard light-profile fitting tools and extracts contiguous features that are significant above the sky background and a noise-only expectation of pixel-contiguity. When applied to both empirical and simulated data, this tool will facilitate future efforts to calibrate the observables associated with galaxy merging signatures and provide improved merger rate constraints.
	
	We apply our residual feature extraction method to single-S\'ersic residual images of $16$ select galaxies from the CANDELS survey \citep[][see Table\,\ref{table_can_test_sample}]{van_der_wel_12} and extract different structures hosted by these galaxies  -- disk structures (4 galaxies), spiral substructures (4 galaxies), and plausible interaction/merger-triggered tidal features (8 galaxies). The main conclusions of our empirical exercise are:
	
	$\bullet$  We extract residual disk features and two-armed spiral substructures for all example galaxies visually identified as hosting disk and spiral features, respectively.
	
	$\bullet$ We extract plausible structurally-similar, dual tidal interaction signatures hosted by {\it three} example close-pair systems, {\it two} close-pair systems where only one galaxy exhibits plausible interaction signatures, {\it one} system experiencing a plausible minor merger event, and {\it one} galaxy hosting plausible post-merger signature.
	
	$\bullet$ To illustrate the role of imaging depth during feature extraction, we compare the residual substructures extracted from deep ({\it HST} $10$-orbit) and shallow ({\it HST} $2$-orbit) images of three example galaxies hosting plausible tidal features. We find that the features extracted at shallow depths are qualitatively similar but noisier in appearance compared to their deeper counterparts owing to the limited sensitivity of shallow CANDELS data to only the higher-surface brightness substructure of the features.
	
	We also apply our tool to synthetic {\it HST} observations of a galaxy merger from the VELA hydro-dynamic simulations \citep{Ceverino14}. We extract the residual features and quantify their areas during $10$ time steps of the merging process spanning $1.7\lesssim z \leq 1.1$ ($\Delta t\sim1.4\,{\rm Gyr}$), each at $19$ viewing orientation angles, and two observational depth configurations matching empirical CANDELS observations (Wide and Deep). We also illustrate the qualitative appearance of these residual features at one fixed orientation. The main conclusions of our theoretical exercise are:
	
	$\bullet$ We find that the features typically rise in {\it prominence} (feature area) $\sim 150\,{\rm Myr}$ after the first pericentric passage ($z=1.63$; time since coalescence $t=-0.46\,{\rm Gyr}$), which visually correspond to close-pair interaction signatures. These features abruptly diminish by a factor of two in significance during $t=-0.15\,{\rm Gyr}$ before coalescence.
	
	$\bullet$ Starting with the coalescence stage ($z=1.44$; $t=0\,{\rm Gyr}$), we extract post-merger features that grow in prominence until peaking at $3-7$ times the initial minimum value after $300\,{\rm Myr}$. These stages visually represent the birth and growing prevalence of tidal arm features. At later time steps $0.47\leq t \leq 0.79\,{\rm Gyr}$, the feature prevalence follows a stochastic rising and falling trend, where the extracted tidal features appear structurally fragmented and clumpy appearance. 
	
	$\bullet$ Imaging depth and viewing orientation play a key role in detecting certain merger features. At deep image depths, we extract all the features at nearly all viewing angles ($\sim 100\%$ detection probability). However,  the same features in shallow depth images are detected at $\sim 50-75\%$ of the viewing sightliness during $-0.46 \leq t \leq -0.15\,{\rm Gyr}$ prior to coalescence. They only reach $100\%$ detection probability at $t>0\,{\rm Gyr}$, when the first noticeable increase in feature prominences occur. Additionally, the temporal trends in shallow depth images are shifted to systematically smaller areas than their deeper depth counterparts. At a fixed depth, the measured feature prominences between different viewing orientations can vary by $20-40\%$.
	
	{While the application of feature extraction method provides important insights about the onset and evolution of tidal features hosted by an example galaxy major merger from the VELA hydro-simulations, we caution the extrapolation of our quantitative (e.g., observability timescales) and qualitative (e.g., tidal feature shapes) interpretations to a general sample of mergers. This is because the strength and appearance of tidal debris may depend on the intrinsic properties of merging galaxies such as: stellar mass-ratio, redshift (driven by the evolution gas-fractions), merger orbital configurations (e.g., prograde vs. retrograde approach). Moreover, it is also important to note that the tidal feature observability carried out in a cosmological context may differ from a similar exercise on idealized binary-merger simulations.}
	
	Lastly, we discuss two additional possible future applications of our general purpose tool. First, we illustrate the extraction of strong gravitational arcs in a candidate strong lensing system identified by \cite{Hocking18}.  Second, we demonstrate the future prospect of using our method to probe high-redshift tidal features by applying it to synthetic {\it JWST} observations of a $z\sim 3$ simulated major merger. We find that {\it JWST} can probe faint, high-redshift tidal features with a smaller exposure time investment and longer rest-frame wavelengths ($2.7\,\mu{\rm m}$) than {\it HST}/WFC3 {\it H}-band $10$-orbit (i.e., CANDELS/Deep) observations.
	
	\section*{Acknowledgments}
	We thank Roberto Abraham for the constructive referee suggestions and feedback that improved this work. We thank Mark Brodwin, Gabriela Canalizo, Alexander De la Vega, Boris H\"au{\ss}ler, Marc Huertas-Company, Kartheik Iyer, Marziye Jafariyazani, Erin Kado-Fong, Allison Kirkpatrick, David Koo, Peter Kurczynski, Bret Lehmer, Jennifer Lotz, Bahram Mobasher, Ripon Saha, and Xianzhong Zheng for their helpful suggestions. KBM, DHM, LDL, and RE acknowledge funding from the NASA Hubble Space Telescope Archival Research grant 15040. CPC, RE, LBF, LDL, and SET acknowledge support from the Missouri Consortium of NASA's National Space Grant College and Fellowship Program. KBM also acknowledges funding from the School of Graduate Studies (SGS) fellowship grant and the Ronald A. MacQuarrie graduate fellowship offered by the University of Missouri-Kansas City.  DC acknowledges funding by the ERC Advanced Grant, STARLIGHT: Formation of the First Stars (project number 339177). This work is based on observations taken by the CANDELS Multi-Cycle Treasury Program with the NASA/ESA HST, which is operated by the Association of Universities for Research in Astronomy, Inc., under NASA contract NAS5-26555. Support for Program number HST-GO-12060 was provided by NASA through a grant from the Space Telescope Science Institute, which is operated by the Association of Universities for Research in Astronomy, Incorporated, under NASA contract NAS5-26555. The VELA simulations were performed at the National Energy Research Scientific Computing Center (NERSC) at Lawrence Berkeley National Laboratory, and at NASA Advanced Supercomputing (NAS) at NASA Ames Research Center. This publication also made use of NASA's Astrophysics Data System Bibliographic Services, TOPCAT \citep[Tools for OPerations on Catalogues And Tables,][]{Taylor05}, the core python package for the astronomy community \citep[{\it Astropy 1.2.1;}][]{Robitaille13}, {\tt collage maker} (\url{github.com/delimitry/collage_maker}).

	\section*{AFFILIATIONS}
	$^1${Department of Physics and Astronomy, University of Missouri-Kansas City, Kansas City, MO 64110, USA.}\\
	$^2${Space Telescope Science Institute, 3700 San Martin Drive, Baltimore, MD 21218, USA.}\\
	$^3${Department of Physics and Astronomy, University of Missouri, Columbia, MO 65211-7010, USA.}\\
	$^4${Department of Physics and Astronomy, Colby College, Waterville, ME 04961, USA.}\\
	$^5${Department of Physics and Astronomy, Johns Hopkins University, Baltimore, MD 21218, USA.}\\
	$^6${Department of Astronomy, University of Michigan, Ann Arbor, MI, 48109-1107, USA.}\\
	$^7${Universit\"at Heidelberg, Zentrum fur Astronomie, Institut f\"ur Theoretische Astrophysik, 69120, Heidelberg, Germany.}\\
	$^8${Physics Department, University of California, Santa Cruz, CA, 95064, USA.}\\
	$^9${Santa Cruz Institute for Particle Physics, University of California, Santa Cruz, CA 95064, USA.}
	\bibliographystyle{mnras} 
	\bibliography{Mantha02_morph_substructure_extraction_paper}

\begin{thebibliography}{}
\makeatletter
\relax
\def\mn@urlcharsother{\let\do\@makeother \do\$\do\&\do\#\do\^\do\_\do\%\do\~}
\def\mn@doi{\begingroup\mn@urlcharsother \@ifnextchar [ {\mn@doi@}
  {\mn@doi@[]}}
\def\mn@doi@[#1]#2{\def\@tempa{#1}\ifx\@tempa\@empty \href
  {http://dx.doi.org/#2} {doi:#2}\else \href {http://dx.doi.org/#2} {#1}\fi
  \endgroup}
\def\mn@eprint#1#2{\mn@eprint@#1:#2::\@nil}
\def\mn@eprint@arXiv#1{\href {http://arxiv.org/abs/#1} {{\tt arXiv:#1}}}
\def\mn@eprint@dblp#1{\href {http://dblp.uni-trier.de/rec/bibtex/#1.xml}
  {dblp:#1}}
\def\mn@eprint@#1:#2:#3:#4\@nil{\def\@tempa {#1}\def\@tempb {#2}\def\@tempc
  {#3}\ifx \@tempc \@empty \let \@tempc \@tempb \let \@tempb \@tempa \fi \ifx
  \@tempb \@empty \def\@tempb {arXiv}\fi \@ifundefined
  {mn@eprint@\@tempb}{\@tempb:\@tempc}{\expandafter \expandafter \csname
  mn@eprint@\@tempb\endcsname \expandafter{\@tempc}}}

\bibitem[\protect\citeauthoryear{{Abraham}, {van den Bergh}, {Glazebrook},
  {Ellis}, {Santiago}, {Surma}  \& {Griffiths}}{{Abraham}
  et~al.}{1996a}]{Abraham96b}
{Abraham} R.~G.,  {van den Bergh} S.,  {Glazebrook} K.,  {Ellis} R.~S.,
  {Santiago} B.~X.,  {Surma} P.,   {Griffiths} R.~E.,  1996a, \mn@doi [The
  Astrophysical Journal Supplement Series] {10.1086/192352}, \href
  {https://ui.adsabs.harvard.edu/\#abs/1996ApJS..107....1A} {107, 1}

\bibitem[\protect\citeauthoryear{{Abraham}, {Tanvir}, {Santiago}, {Ellis},
  {Glazebrook}  \& {van den Bergh}}{{Abraham} et~al.}{1996b}]{Abraham96a}
{Abraham} R.~G.,  {Tanvir} N.~R.,  {Santiago} B.~X.,  {Ellis} R.~S.,
  {Glazebrook} K.,   {van den Bergh} S.,  1996b, \mn@doi [\mnras]
  {10.1093/mnras/279.3.L47}, \href
  {https://ui.adsabs.harvard.edu/\#abs/1996MNRAS.279L..47A} {279, L47}

\bibitem[\protect\citeauthoryear{{Abraham}, {van den Bergh}  \&
  {Nair}}{{Abraham} et~al.}{2003}]{Abraham03}
{Abraham} R.~G.,  {van den Bergh} S.,   {Nair} P.,  2003, \mn@doi [\apj]
  {10.1086/373919}, \href {http://adsabs.harvard.edu/abs/2003ApJ...588..218A}
  {588, 218}

\bibitem[\protect\citeauthoryear{{Astropy Collaboration} et~al.,}{{Astropy
  Collaboration} et~al.}{2013}]{Robitaille13}
{Astropy Collaboration} et~al., 2013, \mn@doi [\aap]
  {10.1051/0004-6361/201322068}, \href
  {http://adsabs.harvard.edu/abs/2013A%26A...558A..33A} {558, A33}

\bibitem[\protect\citeauthoryear{Barbary, Boone, McCully, Craig, Deil  \&
  Rose}{Barbary et~al.}{2016}]{barbary16}
Barbary K.,  Boone K.,  McCully C.,  Craig M.,  Deil C.,   Rose B.,  2016,
  kbarbary/sep: v1.0.0, \mn@doi{10.5281/zenodo.159035}, \url
  {https://doi.org/10.5281/zenodo.159035}

\bibitem[\protect\citeauthoryear{{Barden}, {H{\"a}u{\ss}ler}, {Peng},
  {McIntosh}  \& {Guo}}{{Barden} et~al.}{2012}]{Barden12}
{Barden} M.,  {H{\"a}u{\ss}ler} B.,  {Peng} C.~Y.,  {McIntosh} D.~H.,   {Guo}
  Y.,  2012, \mn@doi [\mnras] {10.1111/j.1365-2966.2012.20619.x}, \href
  {http://adsabs.harvard.edu/abs/2012MNRAS.422..449B} {422, 449}

\bibitem[\protect\citeauthoryear{{Barnes}}{{Barnes}}{1988}]{Barnes88}
{Barnes} J.~E.,  1988, \mn@doi [\apj] {10.1086/166593}, \href
  {http://adsabs.harvard.edu/abs/1988ApJ...331..699B} {331, 699}

\bibitem[\protect\citeauthoryear{{Barnes} \& {Hernquist}}{{Barnes} \&
  {Hernquist}}{1996}]{BarnesHernquist96}
{Barnes} J.~E.,  {Hernquist} L.,  1996, \mn@doi [\apj] {10.1086/177957}, \href
  {http://adsabs.harvard.edu/abs/1996ApJ...471..115B} {471, 115}

\bibitem[\protect\citeauthoryear{Bell et~al.,}{Bell
  et~al.}{2006a}]{bell_dry_2006}
Bell E.~F.,  et~al., 2006a, \mn@doi [ApJ] {10.1086/499931}, 640, 241

\bibitem[\protect\citeauthoryear{{Bell}, {Phleps}, {Somerville}, {Wolf},
  {Borch}  \& {Meisenheimer}}{{Bell} et~al.}{2006b}]{bell06b}
{Bell} E.~F.,  {Phleps} S.,  {Somerville} R.~S.,  {Wolf} C.,  {Borch} A.,
  {Meisenheimer} K.,  2006b, \mn@doi [\apj] {10.1086/508408}, \href
  {http://adsabs.harvard.edu/abs/2006ApJ...652..270B} {652, 270}

\bibitem[\protect\citeauthoryear{{Bertin} \& {Arnouts}}{{Bertin} \&
  {Arnouts}}{1996}]{bertin96}
{Bertin} E.,  {Arnouts} S.,  1996, \mn@doi [\aaps] {10.1051/aas:1996164}, \href
  {http://adsabs.harvard.edu/abs/1996A%26AS..117..393B} {117, 393}

\bibitem[\protect\citeauthoryear{{Bournaud} \& {Duc}}{{Bournaud} \&
  {Duc}}{2006}]{Bournaud06}
{Bournaud} F.,  {Duc} P.-A.,  2006, \mn@doi [\aap]
  {10.1051/0004-6361:20065248}, \href
  {http://adsabs.harvard.edu/abs/2006A%26A...456..481B} {456, 481}

\bibitem[\protect\citeauthoryear{{Brammer} et~al.,}{{Brammer}
  et~al.}{2012}]{Brammer12}
{Brammer} G.~B.,  et~al., 2012, \mn@doi [\apjs] {10.1088/0067-0049/200/2/13},
  \href {http://adsabs.harvard.edu/abs/2012ApJS..200...13B} {200, 13}

\bibitem[\protect\citeauthoryear{{Bruce} et~al.,}{{Bruce}
  et~al.}{2012}]{Bruce12}
{Bruce} V.~A.,  et~al., 2012, \mn@doi [\mnras]
  {10.1111/j.1365-2966.2012.22087.x}, \href
  {http://adsabs.harvard.edu/abs/2012MNRAS.427.1666B} {427, 1666}

\bibitem[\protect\citeauthoryear{{Bruce} et~al.,}{{Bruce}
  et~al.}{2014}]{Bruce14}
{Bruce} V.~A.,  et~al., 2014, \mn@doi [\mnras] {10.1093/mnras/stu1478}, \href
  {http://adsabs.harvard.edu/abs/2014MNRAS.444.1001B} {444, 1001}

\bibitem[\protect\citeauthoryear{{Bruzual} \& {Charlot}}{{Bruzual} \&
  {Charlot}}{2003}]{BC03}
{Bruzual} G.,  {Charlot} S.,  2003, \mn@doi [\mnras]
  {10.1046/j.1365-8711.2003.06897.x}, \href
  {http://adsabs.harvard.edu/abs/2003MNRAS.344.1000B} {344, 1000}

\bibitem[\protect\citeauthoryear{Burger, Burge, Burge  \& Burge}{Burger
  et~al.}{2009}]{Burger09}
Burger W.,  Burge M.~J.,  Burge M.~J.,   Burge M.~J.,  2009, Principles of
  digital image processing.
Springer

\bibitem[\protect\citeauthoryear{{Carlberg}, {Pritchet}  \&
  {Infante}}{{Carlberg} et~al.}{1994}]{Carlberg94}
{Carlberg} R.~G.,  {Pritchet} C.~J.,   {Infante} L.,  1994, \mn@doi [\apj]
  {10.1086/174835}, \href {http://adsabs.harvard.edu/abs/1994ApJ...435..540C}
  {435, 540}

\bibitem[\protect\citeauthoryear{{Cassata} et~al.,}{{Cassata}
  et~al.}{2005}]{cassata05}
{Cassata} P.,  et~al., 2005, \mn@doi [\mnras]
  {10.1111/j.1365-2966.2005.08657.x}, \href
  {http://adsabs.harvard.edu/abs/2005MNRAS.357..903C} {357, 903}

\bibitem[\protect\citeauthoryear{{Ceverino} \& {Klypin}}{{Ceverino} \&
  {Klypin}}{2009}]{Ceverino09}
{Ceverino} D.,  {Klypin} A.,  2009, \mn@doi [\apj]
  {10.1088/0004-637X/695/1/292}, \href
  {http://adsabs.harvard.edu/abs/2009ApJ...695..292C} {695, 292}

\bibitem[\protect\citeauthoryear{{Ceverino}, {Klypin}, {Klimek},
  {Trujillo-Gomez}, {Churchill}, {Primack}  \& {Dekel}}{{Ceverino}
  et~al.}{2014}]{Ceverino14}
{Ceverino} D.,  {Klypin} A.,  {Klimek} E.~S.,  {Trujillo-Gomez} S.,
  {Churchill} C.~W.,  {Primack} J.,   {Dekel} A.,  2014, \mn@doi [\mnras]
  {10.1093/mnras/stu956}, \href
  {http://adsabs.harvard.edu/abs/2014MNRAS.442.1545C} {442, 1545}

\bibitem[\protect\citeauthoryear{{Ceverino}, {Primack}  \& {Dekel}}{{Ceverino}
  et~al.}{2015}]{Ceverino15b}
{Ceverino} D.,  {Primack} J.,   {Dekel} A.,  2015, \mn@doi [\mnras]
  {10.1093/mnras/stv1603}, \href
  {http://adsabs.harvard.edu/abs/2015MNRAS.453..408C} {453, 408}

\bibitem[\protect\citeauthoryear{{Ceverino}, {S{\'a}nchez Almeida}, {Mu{\~n}oz
  Tu{\~n}{\'o}n}, {Dekel}, {Elmegreen}, {Elmegreen}  \& {Primack}}{{Ceverino}
  et~al.}{2016a}]{Ceverino16a}
{Ceverino} D.,  {S{\'a}nchez Almeida} J.,  {Mu{\~n}oz Tu{\~n}{\'o}n} C.,
  {Dekel} A.,  {Elmegreen} B.~G.,  {Elmegreen} D.~M.,   {Primack} J.,  2016a,
  \mn@doi [\mnras] {10.1093/mnras/stw064}, \href
  {http://adsabs.harvard.edu/abs/2016MNRAS.457.2605C} {457, 2605}

\bibitem[\protect\citeauthoryear{{Ceverino}, {Arribas}, {Colina},
  {Rodr{\'{\i}}guez Del Pino}, {Dekel}  \& {Primack}}{{Ceverino}
  et~al.}{2016b}]{Ceverino16b}
{Ceverino} D.,  {Arribas} S.,  {Colina} L.,  {Rodr{\'{\i}}guez Del Pino} B.,
  {Dekel} A.,   {Primack} J.,  2016b, \mn@doi [\mnras] {10.1093/mnras/stw1195},
  \href {http://adsabs.harvard.edu/abs/2016MNRAS.460.2731C} {460, 2731}

\bibitem[\protect\citeauthoryear{{Chabrier}}{{Chabrier}}{2003}]{Chabrier03}
{Chabrier} G.,  2003, \mn@doi [\pasp] {10.1086/376392}, \href
  {http://adsabs.harvard.edu/abs/2003PASP..115..763C} {115, 763}

\bibitem[\protect\citeauthoryear{{Conselice}, {Bershady}, {Dickinson}  \&
  {Papovich}}{{Conselice} et~al.}{2003}]{conselice03a}
{Conselice} C.~J.,  {Bershady} M.~A.,  {Dickinson} M.,   {Papovich} C.,  2003,
  \mn@doi [\aj] {10.1086/377318}, \href
  {http://adsabs.harvard.edu/abs/2003AJ....126.1183C} {126, 1183}

\bibitem[\protect\citeauthoryear{Conselice, Rajgor  \& Myers}{Conselice
  et~al.}{2008}]{conselice_structures_2008}
Conselice C.~J.,  Rajgor S.,   Myers R.,  2008, \mn@doi [Monthly Notices of the
  Royal Astronomical Society] {10.1111/j.1365-2966.2008.13069.x}, 386, 909

\bibitem[\protect\citeauthoryear{Conselice, Yang  \& Bluck}{Conselice
  et~al.}{2009}]{conselice_structures_2009}
Conselice C.~J.,  Yang C.,   Bluck A. F.~L.,  2009, \mn@doi [Monthly Notices of
  the Royal Astronomical Society] {10.1111/j.1365-2966.2009.14396.x}, 394, 1956

\bibitem[\protect\citeauthoryear{Dahlen et~al.,}{Dahlen
  et~al.}{2013}]{dahlen_critical_2013}
Dahlen T.,  et~al., 2013, \mn@doi [ApJ] {10.1088/0004-637X/775/2/93}, 775, 93

\bibitem[\protect\citeauthoryear{{Dimauro} et~al.,}{{Dimauro}
  et~al.}{2018}]{Dimauro18}
{Dimauro} P.,  et~al., 2018, \mn@doi [\mnras] {10.1093/mnras/sty1379}, \href
  {http://adsabs.harvard.edu/abs/2018MNRAS.478.5410D} {478, 5410}

\bibitem[\protect\citeauthoryear{{Draine} \& {Li}}{{Draine} \&
  {Li}}{2007}]{Draine07}
{Draine} B.~T.,  {Li} A.,  2007, \mn@doi [\apj] {10.1086/511055}, \href
  {http://adsabs.harvard.edu/abs/2007ApJ...657..810D} {657, 810}

\bibitem[\protect\citeauthoryear{{Duc} \& {Renaud}}{{Duc} \&
  {Renaud}}{2013}]{Duc2013}
{Duc} P.-A.,  {Renaud} F.,  2013, in {Souchay} J.,  {Mathis} S.,   {Tokieda}
  T.,  eds,  Lecture Notes in Physics, Berlin Springer Verlag Vol. 861, Lecture
  Notes in Physics, Berlin Springer Verlag. p.~327 (\mn@eprint {arXiv}
  {1112.1922}), \mn@doi{10.1007/978-3-642-32961-6_9}

\bibitem[\protect\citeauthoryear{{Dwek}}{{Dwek}}{1998}]{Dwek98}
{Dwek} E.,  1998, \mn@doi [\apj] {10.1086/305829}, \href
  {http://adsabs.harvard.edu/abs/1998ApJ...501..643D} {501, 643}

\bibitem[\protect\citeauthoryear{{Eneev}, {Kozlov}  \& {Sunyaev}}{{Eneev}
  et~al.}{1973}]{Eneev73}
{Eneev} T.~M.,  {Kozlov} N.~N.,   {Sunyaev} R.~A.,  1973, \aap, \href
  {http://adsabs.harvard.edu/abs/1973A%26A....22...41E} {22, 41}

\bibitem[\protect\citeauthoryear{{Fensch} et~al.,}{{Fensch}
  et~al.}{2017}]{Fensch17}
{Fensch} J.,  et~al., 2017, \mn@doi [\mnras] {10.1093/mnras/stw2920}, \href
  {http://adsabs.harvard.edu/abs/2017MNRAS.465.1934F} {465, 1934}

\bibitem[\protect\citeauthoryear{Fiorio \& Gustedt}{Fiorio \&
  Gustedt}{1996}]{Fiorio96}
Fiorio C.,  Gustedt J.,  1996, \mn@doi [Theoretical Computer Science]
  {https://doi.org/10.1016/0304-3975(94)00262-2}, 154, 165

\bibitem[\protect\citeauthoryear{Galametz et~al.,}{Galametz
  et~al.}{2013}]{galametz_candels_2013}
Galametz A.,  et~al., 2013, \mn@doi [ApJ Supplement Series]
  {10.1088/0067-0049/206/2/10}, 206, 10

\bibitem[\protect\citeauthoryear{Grogin et~al.,}{Grogin
  et~al.}{2011}]{grogin_candels:_2011}
Grogin N.~A.,  et~al., 2011, \mn@doi [ApJ Supplement Series]
  {10.1088/0067-0049/197/2/35}, 197, 35

\bibitem[\protect\citeauthoryear{Guo et~al.,}{Guo
  et~al.}{2013}]{guo_candels_2013}
Guo Y.,  et~al., 2013, \mn@doi [ApJ Supplement Series]
  {10.1088/0067-0049/207/2/24}, 207, 24

\bibitem[\protect\citeauthoryear{{Hocking}, {Geach}, {Sun}  \&
  {Davey}}{{Hocking} et~al.}{2018}]{Hocking18}
{Hocking} A.,  {Geach} J.~E.,  {Sun} Y.,   {Davey} N.,  2018, \mn@doi [\mnras]
  {10.1093/mnras/stx2351}, \href
  {http://adsabs.harvard.edu/abs/2018MNRAS.473.1108H} {473, 1108}

\bibitem[\protect\citeauthoryear{{Hopkins}, {Cox}, {Younger}  \&
  {Hernquist}}{{Hopkins} et~al.}{2009}]{Hopkins09}
{Hopkins} P.~F.,  {Cox} T.~J.,  {Younger} J.~D.,   {Hernquist} L.,  2009,
  \mn@doi [\apj] {10.1088/0004-637X/691/2/1168}, \href
  {http://adsabs.harvard.edu/abs/2009ApJ...691.1168H} {691, 1168}

\bibitem[\protect\citeauthoryear{Hopkins et~al.,}{Hopkins
  et~al.}{2010}]{hopkins_mergers_2010}
Hopkins P.~F.,  et~al., 2010, \mn@doi [ApJ] {10.1088/0004-637X/715/1/202}, 715,
  202

\bibitem[\protect\citeauthoryear{{Hoyos} et~al.,}{{Hoyos}
  et~al.}{2012}]{Hoyos12}
{Hoyos} C.,  et~al., 2012, \mn@doi [\mnras] {10.1111/j.1365-2966.2011.19918.x},
  \href {http://adsabs.harvard.edu/abs/2012MNRAS.419.2703H} {419, 2703}

\bibitem[\protect\citeauthoryear{{Hsieh}, {Yee}, {Lin}, {Gladders}  \&
  {Gilbank}}{{Hsieh} et~al.}{2008}]{hsieh08}
{Hsieh} B.~C.,  {Yee} H.~K.~C.,  {Lin} H.,  {Gladders} M.~D.,   {Gilbank}
  D.~G.,  2008, \mn@doi [\apj] {10.1086/589140}, \href
  {http://adsabs.harvard.edu/abs/2008ApJ...683...33H} {683, 33}

\bibitem[\protect\citeauthoryear{{Huertas-Company} et~al.,}{{Huertas-Company}
  et~al.}{2018}]{Huertas-Company18}
{Huertas-Company} M.,  et~al., 2018, preprint, \href
  {http://adsabs.harvard.edu/abs/2018arXiv180407307H} {} (\mn@eprint {arXiv}
  {1804.07307})

\bibitem[\protect\citeauthoryear{{Inoue}, {Dekel}, {Mandelker}, {Ceverino},
  {Bournaud}  \& {Primack}}{{Inoue} et~al.}{2016}]{Inoue16}
{Inoue} S.,  {Dekel} A.,  {Mandelker} N.,  {Ceverino} D.,  {Bournaud} F.,
  {Primack} J.,  2016, \mn@doi [\mnras] {10.1093/mnras/stv2793}, \href
  {http://adsabs.harvard.edu/abs/2016MNRAS.456.2052I} {456, 2052}

\bibitem[\protect\citeauthoryear{{Jiang} et~al.,}{{Jiang}
  et~al.}{2018}]{Jiang18}
{Jiang} F.,  et~al., 2018, preprint, \href
  {http://adsabs.harvard.edu/abs/2018arXiv180407306J} {} (\mn@eprint {arXiv}
  {1804.07306})

\bibitem[\protect\citeauthoryear{{Jogee}}{{Jogee}}{2009}]{Jogee09}
{Jogee} S.,  2009, in {Andersen} J.,  {Nordstr{\"o}ara} {m} B.,
  {Bland-Hawthorn} J.,  eds,  IAU Symposium Vol. 254, The Galaxy Disk in
  Cosmological Context. pp 67--72 (\mn@eprint {arXiv} {0810.5617}),
  \mn@doi{10.1017/S1743921308027403}

\bibitem[\protect\citeauthoryear{{Jonsson}}{{Jonsson}}{2006}]{jonsson06}
{Jonsson} P.,  2006, \mn@doi [\mnras] {10.1111/j.1365-2966.2006.10884.x}, \href
  {http://adsabs.harvard.edu/abs/2006MNRAS.372....2J} {372, 2}

\bibitem[\protect\citeauthoryear{{Jonsson}, {Groves}  \& {Cox}}{{Jonsson}
  et~al.}{2010}]{jonsson10}
{Jonsson} P.,  {Groves} B.~A.,   {Cox} T.~J.,  2010, \mn@doi [\mnras]
  {10.1111/j.1365-2966.2009.16087.x}, \href
  {http://adsabs.harvard.edu/abs/2010MNRAS.403...17J} {403, 17}

\bibitem[\protect\citeauthoryear{{Kado-Fong} et~al.,}{{Kado-Fong}
  et~al.}{2018}]{Kado-Fong18}
{Kado-Fong} E.,  et~al., 2018, preprint, \href
  {http://adsabs.harvard.edu/abs/2018arXiv180505970K} {} (\mn@eprint {arXiv}
  {1805.05970})

\bibitem[\protect\citeauthoryear{Kartaltepe et~al.,}{Kartaltepe
  et~al.}{2007}]{kartaltepe_evolution_2007}
Kartaltepe J.~S.,  et~al., 2007, \mn@doi [ApJ Supplement Series]
  {10.1086/519953}, 172, 320

\bibitem[\protect\citeauthoryear{{Kitzbichler} \& {White}}{{Kitzbichler} \&
  {White}}{2008}]{Kitzbichler_White_08}
{Kitzbichler} M.~G.,  {White} S.~D.~M.,  2008, \mn@doi [\mnras]
  {10.1111/j.1365-2966.2008.13873.x}, \href
  {http://adsabs.harvard.edu/abs/2008MNRAS.391.1489K} {391, 1489}

\bibitem[\protect\citeauthoryear{{Kochanek}}{{Kochanek}}{1995}]{Kochanek95}
{Kochanek} C.~S.,  1995, \mn@doi [\apj] {10.1086/175721}, \href
  {http://adsabs.harvard.edu/abs/1995ApJ...445..559K} {445, 559}

\bibitem[\protect\citeauthoryear{{Koekemoer} et~al.,}{{Koekemoer}
  et~al.}{2011}]{Koekemoer}
{Koekemoer} A.~M.,  et~al., 2011, \mn@doi [\apjs] {10.1088/0067-0049/197/2/36},
  \href {http://adsabs.harvard.edu/abs/2011ApJS..197...36K} {197, 36}

\bibitem[\protect\citeauthoryear{{Kravtsov}}{{Kravtsov}}{2003}]{kravtsov03}
{Kravtsov} A.~V.,  2003, \mn@doi [\apjl] {10.1086/376674}, \href
  {http://adsabs.harvard.edu/abs/2003ApJ...590L...1K} {590, L1}

\bibitem[\protect\citeauthoryear{{Kravtsov}, {Klypin}  \&
  {Khokhlov}}{{Kravtsov} et~al.}{1997}]{kravtsov97}
{Kravtsov} A.~V.,  {Klypin} A.~A.,   {Khokhlov} A.~M.,  1997, \mn@doi [\apjs]
  {10.1086/313015}, \href {http://adsabs.harvard.edu/abs/1997ApJS..111...73K}
  {111, 73}

\bibitem[\protect\citeauthoryear{{Krist}, {Hook}  \& {Stoehr}}{{Krist}
  et~al.}{2011}]{Krist11}
{Krist} J.~E.,  {Hook} R.~N.,   {Stoehr} F.,  2011, in Optical Modeling and
  Performance Predictions V. p. 81270J, \mn@doi{10.1117/12.892762}

\bibitem[\protect\citeauthoryear{{Lackner} et~al.,}{{Lackner}
  et~al.}{2014}]{Lackner14}
{Lackner} C.~N.,  et~al., 2014, \mn@doi [\aj] {10.1088/0004-6256/148/6/137},
  \href {http://adsabs.harvard.edu/abs/2014AJ....148..137L} {148, 137}

\bibitem[\protect\citeauthoryear{{Lin} et~al.,}{{Lin} et~al.}{2004}]{Lin04}
{Lin} L.,  et~al., 2004, \mn@doi [\apjl] {10.1086/427183}, \href
  {http://adsabs.harvard.edu/abs/2004ApJ...617L...9L} {617, L9}

\bibitem[\protect\citeauthoryear{Lin et~al.,}{Lin
  et~al.}{2008}]{lin_redshift_2008}
Lin L.,  et~al., 2008, \mn@doi [ApJ] {10.1086/587928}, 681, 232

\bibitem[\protect\citeauthoryear{{L{\'o}pez-Sanjuan}, {Balcells},
  {P{\'e}rez-Gonz{\'a}lez}, {Barro}, {Garc{\'{\i}}a-Dab{\'o}}, {Gallego}  \&
  {Zamorano}}{{L{\'o}pez-Sanjuan} et~al.}{2009}]{Lopez-Sanjuan09}
{L{\'o}pez-Sanjuan} C.,  {Balcells} M.,  {P{\'e}rez-Gonz{\'a}lez} P.~G.,
  {Barro} G.,  {Garc{\'{\i}}a-Dab{\'o}} C.~E.,  {Gallego} J.,   {Zamorano} J.,
  2009, \mn@doi [\aap] {10.1051/0004-6361/200911923}, \href
  {http://adsabs.harvard.edu/abs/2009A%26A...501..505L} {501, 505}

\bibitem[\protect\citeauthoryear{{Lotz} et~al.,}{{Lotz} et~al.}{2008}]{lotz08}
{Lotz} J.~M.,  et~al., 2008, \mn@doi [\apj] {10.1086/523659}, \href
  {http://adsabs.harvard.edu/abs/2008ApJ...672..177L} {672, 177}

\bibitem[\protect\citeauthoryear{Lotz, Jonsson, Cox  \& Primack}{Lotz
  et~al.}{2010}]{lotz_effect_2010}
Lotz J.~M.,  Jonsson P.,  Cox T.~J.,   Primack J.~R.,  2010, \mn@doi [Monthly
  Notices of the Royal Astronomical Society]
  {10.1111/j.1365-2966.2010.16268.x}, 404, 575

\bibitem[\protect\citeauthoryear{Lotz, Jonsson, Cox, Croton, Primack,
  Somerville  \& Stewart}{Lotz et~al.}{2011}]{lotz_major_2011}
Lotz J.~M.,  Jonsson P.,  Cox T.~J.,  Croton D.,  Primack J.~R.,  Somerville
  R.~S.,   Stewart K.,  2011, \mn@doi [ApJ] {10.1088/0004-637X/742/2/103}, 742,
  103

\bibitem[\protect\citeauthoryear{{Man}, {Zirm}  \& {Toft}}{{Man}
  et~al.}{2016}]{man16}
{Man} A.~W.~S.,  {Zirm} A.~W.,   {Toft} S.,  2016, \mn@doi [\apj]
  {10.3847/0004-637X/830/2/89}, \href
  {http://adsabs.harvard.edu/abs/2016ApJ...830...89M} {830, 89}

\bibitem[\protect\citeauthoryear{{Mandelker}, {Dekel}, {Ceverino}, {DeGraf},
  {Guo}  \& {Primack}}{{Mandelker} et~al.}{2017}]{Mandelker17}
{Mandelker} N.,  {Dekel} A.,  {Ceverino} D.,  {DeGraf} C.,  {Guo} Y.,
  {Primack} J.,  2017, \mn@doi [\mnras] {10.1093/mnras/stw2358}, \href
  {http://adsabs.harvard.edu/abs/2017MNRAS.464..635M} {464, 635}

\bibitem[\protect\citeauthoryear{{Mantha} et~al.,}{{Mantha}
  et~al.}{2018}]{Mantha18}
{Mantha} K.~B.,  et~al., 2018, \mn@doi [\mnras] {10.1093/mnras/stx3260}, \href
  {http://adsabs.harvard.edu/abs/2018MNRAS.475.1549M} {475, 1549}

\bibitem[\protect\citeauthoryear{{McIntosh}, {Guo}, {Hertzberg}, {Katz}, {Mo},
  {van den Bosch}  \& {Yang}}{{McIntosh} et~al.}{2008}]{mcintosh_2008}
{McIntosh} D.~H.,  {Guo} Y.,  {Hertzberg} J.,  {Katz} N.,  {Mo} H.~J.,  {van
  den Bosch} F.~C.,   {Yang} X.,  2008, \mn@doi [\mnras]
  {10.1111/j.1365-2966.2008.13531.x}, \href
  {http://adsabs.harvard.edu/abs/2008MNRAS.388.1537M} {388, 1537}

\bibitem[\protect\citeauthoryear{{Mobasher} et~al.,}{{Mobasher}
  et~al.}{2015}]{Mobasher15}
{Mobasher} B.,  et~al., 2015, \mn@doi [\apj] {10.1088/0004-637X/808/1/101},
  \href {http://adsabs.harvard.edu/abs/2015ApJ...808..101M} {808, 101}

\bibitem[\protect\citeauthoryear{{Moody}, {Guo}, {Mandelker}, {Ceverino},
  {Mozena}, {Koo}, {Dekel}  \& {Primack}}{{Moody} et~al.}{2014}]{Moody14}
{Moody} C.~E.,  {Guo} Y.,  {Mandelker} N.,  {Ceverino} D.,  {Mozena} M.,  {Koo}
  D.~C.,  {Dekel} A.,   {Primack} J.,  2014, \mn@doi [\mnras]
  {10.1093/mnras/stu1534}, \href
  {http://adsabs.harvard.edu/abs/2014MNRAS.444.1389M} {444, 1389}

\bibitem[\protect\citeauthoryear{{Mundy}, {Conselice}, {Duncan}, {Almaini},
  {H{\"a}u{\ss}ler}  \& {Hartley}}{{Mundy} et~al.}{2017}]{Mundy17}
{Mundy} C.~J.,  {Conselice} C.~J.,  {Duncan} K.~J.,  {Almaini} O.,
  {H{\"a}u{\ss}ler} B.,   {Hartley} W.~G.,  2017, preprint, \href
  {http://adsabs.harvard.edu/abs/2017arXiv170507986M} {} (\mn@eprint {arXiv}
  {1705.07986})

\bibitem[\protect\citeauthoryear{{Namboodiri} \& {Kochhar}}{{Namboodiri} \&
  {Kochhar}}{1985}]{Namboodiri85}
{Namboodiri} P.~M.~S.,  {Kochhar} R.~K.,  1985, Bulletin of the Astronomical
  Society of India, \href {http://adsabs.harvard.edu/abs/1985BASI...13..363N}
  {13, 363}

\bibitem[\protect\citeauthoryear{Newman, Ellis, Bundy  \& Treu}{Newman
  et~al.}{2012}]{newman_can_2012}
Newman A.~B.,  Ellis R.~S.,  Bundy K.,   Treu T.,  2012, \mn@doi [ApJ]
  {10.1088/0004-637X/746/2/162}, 746, 162

\bibitem[\protect\citeauthoryear{{Oke} \& {Gunn}}{{Oke} \&
  {Gunn}}{1983}]{oke83}
{Oke} J.~B.,  {Gunn} J.~E.,  1983, \mn@doi [\apj] {10.1086/160817}, \href
  {http://adsabs.harvard.edu/abs/1983ApJ...266..713O} {266, 713}

\bibitem[\protect\citeauthoryear{{Patton}, {Pritchet}, {Yee}, {Ellingson}  \&
  {Carlberg}}{{Patton} et~al.}{1997}]{Patton97}
{Patton} D.~R.,  {Pritchet} C.~J.,  {Yee} H.~K.~C.,  {Ellingson} E.,
  {Carlberg} R.~G.,  1997, \apj, \href
  {http://adsabs.harvard.edu/abs/1997ApJ...475...29P} {475, 29}

\bibitem[\protect\citeauthoryear{{Patton}, {Carlberg}, {Marzke}, {Pritchet},
  {da Costa}  \& {Pellegrini}}{{Patton} et~al.}{2000}]{Patton00}
{Patton} D.~R.,  {Carlberg} R.~G.,  {Marzke} R.~O.,  {Pritchet} C.~J.,  {da
  Costa} L.~N.,   {Pellegrini} P.~S.,  2000, \mn@doi [\apj] {10.1086/308907},
  \href {http://adsabs.harvard.edu/abs/2000ApJ...536..153P} {536, 153}

\bibitem[\protect\citeauthoryear{{Peirani}, {Crockett}, {Geen}, {Khochfar},
  {Kaviraj}  \& {Silk}}{{Peirani} et~al.}{2010}]{Peirani10}
{Peirani} S.,  {Crockett} R.~M.,  {Geen} S.,  {Khochfar} S.,  {Kaviraj} S.,
  {Silk} J.,  2010, \mn@doi [\mnras] {10.1111/j.1365-2966.2010.16666.x}, \href
  {http://adsabs.harvard.edu/abs/2010MNRAS.405.2327P} {405, 2327}

\bibitem[\protect\citeauthoryear{{Peng}, {Ho}, {Impey}  \& {Rix}}{{Peng}
  et~al.}{2002}]{Peng02}
{Peng} C.~Y.,  {Ho} L.~C.,  {Impey} C.~D.,   {Rix} H.-W.,  2002, \mn@doi [\aj]
  {10.1086/340952}, \href {http://adsabs.harvard.edu/abs/2002AJ....124..266P}
  {124, 266}

\bibitem[\protect\citeauthoryear{{Planck Collaboration} et~al.,}{{Planck
  Collaboration} et~al.}{2016}]{ade15}
{Planck Collaboration} et~al., 2016, \mn@doi [\aap]
  {10.1051/0004-6361/201525830}, \href
  {http://adsabs.harvard.edu/abs/2016A%26A...594A..13P} {594, A13}

\bibitem[\protect\citeauthoryear{{Refsdal}}{{Refsdal}}{1964}]{Refsdal64}
{Refsdal} S.,  1964, \mn@doi [\mnras] {10.1093/mnras/128.4.307}, \href
  {http://adsabs.harvard.edu/abs/1964MNRAS.128..307R} {128, 307}

\bibitem[\protect\citeauthoryear{Reiss}{Reiss}{1993}]{Reiss93}
Reiss T.~H.,  1993, Recognizing planar objects using invariant image features.
Springer-Verlag New York, Inc.

\bibitem[\protect\citeauthoryear{Robaina, Bell, van~der Wel, Somerville,
  Skelton, McIntosh, Meisenheimer  \& Wolf}{Robaina
  et~al.}{2010}]{robaina_merger-driven_2010}
Robaina A.~R.,  Bell E.~F.,  van~der Wel A.,  Somerville R.~S.,  Skelton R.~E.,
   McIntosh D.~H.,  Meisenheimer K.,   Wolf C.,  2010, \mn@doi [ApJ]
  {10.1088/0004-637X/719/1/844}, 719, 844

\bibitem[\protect\citeauthoryear{{Robotham} et~al.,}{{Robotham}
  et~al.}{2014}]{Robotham14}
{Robotham} A.~S.~G.,  et~al., 2014, \mn@doi [\mnras] {10.1093/mnras/stu1604},
  \href {http://adsabs.harvard.edu/abs/2014MNRAS.444.3986R} {444, 3986}

\bibitem[\protect\citeauthoryear{Rodriguez-Gomez et~al.,}{Rodriguez-Gomez
  et~al.}{2015}]{rodriguez-gomez_merger_2015}
Rodriguez-Gomez V.,  et~al., 2015, \mn@doi [Monthly Notices of the Royal
  Astronomical Society] {10.1093/mnras/stv264}, 449, 49

\bibitem[\protect\citeauthoryear{{Ryan}, {Cohen}, {Windhorst}  \&
  {Silk}}{{Ryan} et~al.}{2008}]{Ryan08}
{Ryan} Jr. R.~E.,  {Cohen} S.~H.,  {Windhorst} R.~A.,   {Silk} J.,  2008,
  \mn@doi [\apj] {10.1086/527463}, \href
  {http://adsabs.harvard.edu/abs/2008ApJ...678..751R} {678, 751}

\bibitem[\protect\citeauthoryear{Santini et~al.,}{Santini
  et~al.}{2015}]{santini_stellar_2015}
Santini P.,  et~al., 2015, \mn@doi [ApJ] {10.1088/0004-637X/801/2/97}, 801, 97

\bibitem[\protect\citeauthoryear{{Simons} et~al.,}{{Simons}
  et~al.}{2019}]{Simons19}
{Simons} R.~C.,  et~al., 2019, arXiv e-prints, \href
  {https://ui.adsabs.harvard.edu/\#abs/2019arXiv190206762S} {p.
  arXiv:1902.06762}

\bibitem[\protect\citeauthoryear{{Snyder} et~al.,}{{Snyder}
  et~al.}{2015}]{snyder15}
{Snyder} G.~F.,  et~al., 2015, \mn@doi [\mnras] {10.1093/mnras/stv2078}, \href
  {http://adsabs.harvard.edu/abs/2015MNRAS.454.1886S} {454, 1886}

\bibitem[\protect\citeauthoryear{{Snyder}, {Lotz}, {Rodriguez-Gomez},
  {Guimar{\~a}es}, {Torrey}  \& {Hernquist}}{{Snyder} et~al.}{2017}]{snyder17}
{Snyder} G.~F.,  {Lotz} J.~M.,  {Rodriguez-Gomez} V.,  {Guimar{\~a}es}
  R.~d.~S.,  {Torrey} P.,   {Hernquist} L.,  2017, \mn@doi [\mnras]
  {10.1093/mnras/stx487}, \href
  {http://adsabs.harvard.edu/abs/2017MNRAS.468..207S} {468, 207}

\bibitem[\protect\citeauthoryear{{Tacchella}, {Dekel}, {Carollo}, {Ceverino},
  {DeGraf}, {Lapiner}, {Mandelker}  \& {Primack}}{{Tacchella}
  et~al.}{2016}]{Tacchella16}
{Tacchella} S.,  {Dekel} A.,  {Carollo} C.~M.,  {Ceverino} D.,  {DeGraf} C.,
  {Lapiner} S.,  {Mandelker} N.,   {Primack} J.~R.,  2016, \mn@doi [\mnras]
  {10.1093/mnras/stw303}, \href
  {http://adsabs.harvard.edu/abs/2016MNRAS.458..242T} {458, 242}

\bibitem[\protect\citeauthoryear{{Tal}, {van Dokkum}, {Nelan}  \&
  {Bezanson}}{{Tal} et~al.}{2009}]{Tal09}
{Tal} T.,  {van Dokkum} P.~G.,  {Nelan} J.,   {Bezanson} R.,  2009, \mn@doi
  [\aj] {10.1088/0004-6256/138/5/1417}, \href
  {http://adsabs.harvard.edu/abs/2009AJ....138.1417T} {138, 1417}

\bibitem[\protect\citeauthoryear{{Taylor}}{{Taylor}}{2005}]{Taylor05}
{Taylor} M.~B.,  2005, in {Shopbell} P.,  {Britton} M.,   {Ebert} R.,  eds,
  Astronomical Society of the Pacific Conference Series Vol. 347, Astronomical
  Data Analysis Software and Systems XIV. p.~29

\bibitem[\protect\citeauthoryear{{Tomassetti} et~al.,}{{Tomassetti}
  et~al.}{2016}]{Tomassetti16}
{Tomassetti} M.,  et~al., 2016, \mn@doi [\mnras] {10.1093/mnras/stw606}, \href
  {http://adsabs.harvard.edu/abs/2016MNRAS.458.4477T} {458, 4477}

\bibitem[\protect\citeauthoryear{{Toomre}}{{Toomre}}{1977}]{Toomre77}
{Toomre} A.,  1977, in {Tinsley} B.~M.,  {Larson} D.~Campbell R.~B.~G.,  eds,
  Evolution of Galaxies and Stellar Populations. p.~401

\bibitem[\protect\citeauthoryear{{Toomre} \& {Toomre}}{{Toomre} \&
  {Toomre}}{1972}]{Toomre72}
{Toomre} A.,  {Toomre} J.,  1972, \mn@doi [\apj] {10.1086/151823}, \href
  {http://adsabs.harvard.edu/abs/1972ApJ...178..623T} {178, 623}

\bibitem[\protect\citeauthoryear{{Treu} \& {Koopmans}}{{Treu} \&
  {Koopmans}}{2004}]{Treu04}
{Treu} T.,  {Koopmans} L.~V.~E.,  2004, \mn@doi [\apj] {10.1086/422245}, \href
  {http://adsabs.harvard.edu/abs/2004ApJ...611..739T} {611, 739}

\bibitem[\protect\citeauthoryear{{Tuccillo}, {Huertas-Company},
  {Decenci{\`e}re}, {Velasco-Forero}, {Dom{\'{\i}}nguez S{\'a}nchez}  \&
  {Dimauro}}{{Tuccillo} et~al.}{2018}]{Tuccillo18}
{Tuccillo} D.,  {Huertas-Company} M.,  {Decenci{\`e}re} E.,  {Velasco-Forero}
  S.,  {Dom{\'{\i}}nguez S{\'a}nchez} H.,   {Dimauro} P.,  2018, \mn@doi
  [\mnras] {10.1093/mnras/stx3186}, \href
  {http://adsabs.harvard.edu/abs/2018MNRAS.475..894T} {475, 894}

\bibitem[\protect\citeauthoryear{{Ventou} et~al.,}{{Ventou}
  et~al.}{2017}]{Ventou17}
{Ventou} E.,  et~al., 2017, \mn@doi [\aap] {10.1051/0004-6361/201731586}, \href
  {http://adsabs.harvard.edu/abs/2017A%26A...608A...9V} {608, A9}

\bibitem[\protect\citeauthoryear{{Weingartner} \& {Draine}}{{Weingartner} \&
  {Draine}}{2001}]{Weingartner01}
{Weingartner} J.~C.,  {Draine} B.~T.,  2001, \mn@doi [\apj] {10.1086/318651},
  \href {http://adsabs.harvard.edu/abs/2001ApJ...548..296W} {548, 296}

\bibitem[\protect\citeauthoryear{{Wolf} et~al.,}{{Wolf} et~al.}{2005}]{Wolf05}
{Wolf} C.,  et~al., 2005, \mn@doi [\apj] {10.1086/431659}, \href
  {http://adsabs.harvard.edu/abs/2005ApJ...630..771W} {630, 771}

\bibitem[\protect\citeauthoryear{Wu, Otoo  \& Shoshani}{Wu et~al.}{2005}]{Wu05}
Wu K.,  Otoo E.,   Shoshani A.,  2005, in Medical Imaging 2005: Image
  Processing. pp 1965--1977

\bibitem[\protect\citeauthoryear{{Wuyts} et~al.,}{{Wuyts}
  et~al.}{2013}]{Wuyts13}
{Wuyts} S.,  et~al., 2013, \mn@doi [\apj] {10.1088/0004-637X/779/2/135}, \href
  {http://adsabs.harvard.edu/abs/2013ApJ...779..135W} {779, 135}

\bibitem[\protect\citeauthoryear{{Zepf} \& {Koo}}{{Zepf} \&
  {Koo}}{1989}]{Zepf89}
{Zepf} S.~E.,  {Koo} D.~C.,  1989, \mn@doi [\apj] {10.1086/167085}, \href
  {http://adsabs.harvard.edu/abs/1989ApJ...337...34Z} {337, 34}

\bibitem[\protect\citeauthoryear{{Zolotov} et~al.,}{{Zolotov}
  et~al.}{2015}]{zolotov15}
{Zolotov} A.,  et~al., 2015, \mn@doi [\mnras] {10.1093/mnras/stv740}, \href
  {http://adsabs.harvard.edu/abs/2015MNRAS.450.2327Z} {450, 2327}

\bibitem[\protect\citeauthoryear{{de Ravel} et~al.,}{{de Ravel}
  et~al.}{2009}]{de_ravel09}
{de Ravel} L.,  et~al., 2009, \mn@doi [\aap] {10.1051/0004-6361/200810569},
  \href {http://adsabs.harvard.edu/abs/2009A%26A...498..379D} {498, 379}

\bibitem[\protect\citeauthoryear{de Ravel et~al.,}{de~Ravel
  et~al.}{2011}]{de_ravel_zcosmos_2011}
de Ravel L.,  et~al., 2011, arXiv:1104.5470 [astro-ph]

\bibitem[\protect\citeauthoryear{{van der Wel} et~al.,}{{van der Wel}
  et~al.}{2012}]{van_der_wel_12}
{van der Wel} A.,  et~al., 2012, \mn@doi [\apjs] {10.1088/0067-0049/203/2/24},
  \href {http://adsabs.harvard.edu/abs/2012ApJS..203...24V} {203, 24}

\bibitem[\protect\citeauthoryear{{van der Wel} et~al.,}{{van der Wel}
  et~al.}{2013}]{van_der_Wel13}
{van der Wel} A.,  et~al., 2013, \mn@doi [\apjl] {10.1088/2041-8205/777/1/L17},
  \href {http://adsabs.harvard.edu/abs/2013ApJ...777L..17V} {777, L17}

\bibitem[\protect\citeauthoryear{van~der Wel et~al.,}{van~der Wel
  et~al.}{2014}]{van_der_wel_3d-hst+candels:_2014}
van~der Wel A.,  et~al., 2014, \mn@doi [ApJ] {10.1088/0004-637X/788/1/28}, 788,
  28

\makeatother
\end{thebibliography}
\end{document}